\documentclass[twocolumn]{aastex6}

\usepackage{graphicx}
\usepackage{array}
\usepackage{amssymb}
\usepackage{natbib}
\usepackage{url}
\usepackage[utf8]{inputenc}  
\usepackage{amsmath}  
\usepackage{footnote}
\shorttitle{SAMP II.}
\shortauthors{Rakshit et al.}

\begin{document}


\title{THE SEOUL NATIONAL UNIVERSITY AGN MONITORING PROJECT. II. BLR size and black hole mass of two AGNs}



\author{Suvendu Rakshit$^{1,2}$}
\author{Jong-Hak Woo$^{1}$}
\author{Elena Gallo$^{3}$}
\author{Edmund Hodges-Kluck$^{3,4}$}
\author{Jaejin Shin$^{1}$}
\author{Yiseul Jeon$^{1}$}
\author{Hyun-Jin Bae$^{1}$}
\author{Vivienne Baldassare$^{3}$}
\author{Hojin Cho$^{1}$}
\author{Wanjin Cho$^{1}$}
\author{Adi Foord$^{3}$}
\author{Daeun Kang$^{1}$}
\author{Wonseok Kang$^{7}$}
\author{Marios Karouzos$^{1}$}
\author{Minjin Kim$^{5}$}
\author{Taewoo Kim$^{7}$}
\author{Huynh Anh N. Le$^{1}$}
\author{Daeseong Park$^{6}$}
\author{Songyoun Park$^{1}$}
\author{Donghoon Son$^{1}$}
\author{Hyun-il Sung$^{6}$}
\author{Vardha N. Bennert$^{8}$}
\author{Matthew A. Malkan$^{9}$}

\affil{$^{1}$Astronomy Program, Department of Physics and Astronomy, Seoul National University, Seoul, 08826, Republic of Korea; suvenduat@gmail.com\\
$^{2}$Finnish Centre for Astronomy with ESO (FINCA), University of Turku, Quantum, Vesilinnantie 5, 20014 University of Turku, Finland\\
$^{3}$Department of Astronomy, University of Michigan, Ann Arbor, MI 48109, USA\\
$^{4}$NASA/GSFC, Code 662, Greenbelt, MD 20771, USA
$^{5}$Department of Astronomy and Atmospheric Sciences, Kyungpook National University, Daegu 41566 Korea
$^{6}$Korea Astronomy and Space science Institute, Daejeon, Republic of Korea\\
$^{7}$National Youth Space Center, Goheung, Jeollanam-do, 59567, Korea 
$^{8}$Physics Department, California Polytechnic State University, San Luis Obispo, CA 93407, USA
$^{9}$Department of Physics and Astronomy, University of California, Los Angeles, CA 90095, USA
}

\begin{abstract}
Active galactic nuclei (AGNs) show a correlation between the size of the broad line region (BLR) and the monochromatic continuum luminosity at 5100\AA, allowing black hole mass estimation based on single-epoch spectra. However, the validity of the correlation is yet to be clearly tested for high-luminosity AGNs. We present the first reverberation-mapping results of the Seoul National University AGN monitoring program (SAMP), which is designed to focus on luminous AGNs for probing the high end of the size-luminosity relation. We report time lag measurements of two AGNs, namely, 2MASS J10261389+5237510 and SDSS J161911.24+501109.2, 
using the light curves obtained over a $\sim$1000 day period with an average cadence of $\sim$10 and $\sim$20 days, respectively for photometry and spectroscopy monitoring. Based on a cross-correlation analysis and H$\beta$ line width measurements, we determine the H$\beta$ lag as $41.8^{+4.9}_{-6.0}$ and $52.6^{+17.6}_{-14.7}$ days in the observed-frame, and black hole mass as  $3.65^{+0.49}_{-0.57} \times 10^7 M_{\odot}$ and $23.02^{+7.81}_{-6.56} \times 10^7 M_{\odot}$, respectively for 2MASS J1026 and SDSS J1619. 
\end{abstract}






\section{Introduction}

The correlation between black hole mass and host galaxy properties suggests the connection between black hole growth and galaxy evolution \citep{2013ARA&A..51..511K}. Accurate measurements of black hole masses are required to further study black hole-galaxy co-evolution. For nearby galaxies, black hole mass can be dynamically measured using spatially resolved kinematics of stars \citep[e.g.,][]{1994MNRAS.270..271V} or gas \citep[e.g.,][]{1994ApJ...435L..35H}. In contrast, dynamical measurements are challenging for active galactic nuclei (AGNs) beyond the local volume as an extremely high spatial resolution is required. 

Alternatively, the reverberation mapping technique provides reliable mass measurements for AGNs \citep{1982ApJ...255..419B,1993PASP..105..247P}. 
The time lag between the continuum flux and broad emission line flux variations reflects the size of the broad line region (BLR).
Assuming that the gas motion in the BLR is governed by the black hole's gravitational potential, black hole mass can be determined by combining the size of the BLR ($R_{\mathrm{BLR}}$) and the velocity width of broad emission lines ($\Delta V$) based on the virial relation,
\begin{equation}
M_{\mathrm{BH}}= f \times R_{\mathrm{BLR}}~ \Delta V^2/G 
\label{eq:mass}
\end{equation} 
where $f$ is a dimensionless scale factor that depends on the geometry and kinematics of the BLR.


 Previous reverberation mapping studies provided reliable time-lag measurements for $\sim$100 sources \citep[e.g.,][]{1997ApJ...490L.131W,2000ApJ...533..631K,1998ApJ...501...82P,2002ApJ...581..197P,2004ApJ...613..682P,2009ApJ...705..199B,2010ApJ...716..993B,2013ApJ...767..149B,2009ApJ...704L..80D,2011ApJ...743L...4B,2015ApJS..217...26B,2013ApJ...773...24R,2016ApJ...818...30S,2017ApJ...847..125P,2012ApJ...755...60G,2017ApJ...851...21G,2017ApJ...840...97F,2014ApJ...782...45D,2015ApJ...806...22D,2018ApJ...856....6D,2019ApJ...876...49Z,2019NatAs...3..755W}. These measurements showed that the H$\beta$ BLR size correlates with the monochromatic luminosity at 5100\AA\ \citep{2013ApJ...767..149B}. 
Thus, the black hole mass can be indirectly estimated based on single-epoch spectra by using the monochromatic luminosity at 5100\AA\ as a proxy for $R_{\mathrm{BLR}}$. 
While this method requires a well-calibrated BLR size-luminosity relation, recent reverberation mapping results reported a large scatter of the BLR size-luminosity relation by including high Eddington ratio AGNs \citep{2016ApJ...825..126D,2018ApJ...856....6D} or by investigating single object over time \citep{Pei_2017}.

The reverberation-mapped AGNs are mainly limited to low-to-intermediate luminosity AGNs at relatively low-z. 
These limitations may bias the BLR size-luminosity relation at the high-luminosity end, which is more relevant for estimating black hole masses of luminous high-z QSOs. While the reverberation sample size increased by recent new programs \citep[e.g.,][]{2015ApJS..216....4S,2017ApJ...851...21G}, the moderate-to-high luminosity end of the size-luminosity relation is yet to be clearly explored. There are various reasons for this limitation. First, high-luminosity AGNs are more challenging to study since they are less variable \citep[see][and references therein]{2017ApJ...842...96R}. Second, the time lag is expected to be much longer for more luminous AGNs, requiring a longer time baseline for monitoring. Third, while several studies focused on a sample of luminous high-z AGNs, the optical spectrograph used in the monitoring only probed the rest-frame UV emission lines due to the redshift effect \citep[e.g.,][]{2007ApJ...659..997K,2018ApJ...865...56L}. Consequently, the high luminosity end of the H$\beta$ size-luminosity relation is still virtually unexplored. 

To expand the size-luminosity relation towards high-luminosity and high-mass AGNs, we started the Seoul National University AGN monitoring Project (SAMP), which targets the H$\beta$ lag measurements for luminous AGNs with $L_{5100} \sim 10^{44-46} \, \mathrm{erg \, s^{-1}}$ at $z<0.5$. The project strategy and sample is outlined by \citet{2019arXiv190700771W}. The initial test observations of $\sim$~100 AGNs started in October 2015 with 1-2 m class telescopes in order to determine the magnitude and luminosity of each target and the H$\beta$ and [O III] emission line strengths for feasibility.  Then, the spectroscopic monitoring with the Lick 3-m started in Feb. 2016 with $\sim20$ days cadence, while the photometric cadence was $\sim10$ days. The initial three year campaign along with the 20 days time resolution is suitable to obtain good quality light curves to constrain the expected lag $\sim 100-200$ days \citep[see][]{2019arXiv190700771W}. In this paper, we report the first result of the spectroscopic lag measurements for two targets using the spectroscopic and photometric data obtained over the three year period. 
In section \ref{sec:data} we present the observation and data reduction technique and in section \ref{sec:analysis} we perform the spectral decomposition, time lag, and black hole mass estimation. The results are discussed in section \ref{sec:discussion} and summarized in section \ref{sec:summary}.

\section{Observation and data reduction}\label{sec:data}

\subsection{Targets}

Based on the initial variability test for $\sim$100 AGNs selected by \citet{2019arXiv190700771W}, we further chose 48 main targets, which have the expected lag between 70 and 300 days, for the spectroscopy monitoring. A dozen AGNs shows clear variability in the continuum and H$\beta$ light curves. Here, we present the first 2 AGNs with reliable lag measurements. Note that these lags are relatively short compared to the expected lags of the sample. While at the end of the monitoring campaign we will obtain a number of objects having longer lags, at the moment we present clear measurements for these two AGNs. Monitoring observations of other targets are on-going and the results will be presented in forthcoming papers. We summarize the main properties of the two AGNs.

~

(I) 2MASS J10261389+5237510 at $z=0.259$ with the $B$= 17.96 \citep{2010A&A...518A..10V}. Using the SDSS DR7 spectra, \citet{2011ApJS..194...45S} reported the monochromatic luminosity at 5100\AA\ $\log L_{5100}$ =  $44.35\pm0.01 \,\mathrm{erg \, s^{-1}}$, H$\beta$ emission line FWHM = $3175 \pm 91 \, \mathrm{km s^{-1}}$, weak Fe II emission (with a Fe II/H$\beta$ ratio $R_{\rm Fe II} \sim 0.03$), $\log M_{\mathrm{BH}}$ = $8.09 \pm 0.02 \, M_{\odot}$ and Eddington ratio ($\log \lambda_{\mathrm{EDD}}$) = $-0.86$ for the source. 

(II) SDSS J161911.24+501109.2 at $z=0.234$ with the $B$= 18.05 \citep{2010A&A...518A..10V}. The source has $\log L_{5100} = 44.37\pm0.01 \, \mathrm{erg \, s^{-1}}$, H$\beta~ \rm FWHM$=$4411 \pm 65 \, \mathrm{km s^{-1}}$, $R_{\mathrm{FeII}} \sim 0.24$, $\log M_{\mathrm{BH}}$ = $8.38 \pm 0.01 \, M_{\odot}$ and $\log \lambda_{\mathrm{EDD}} = -1.14$ \citep{2011ApJS..194...45S}.    

\begin{figure*}
\centering
\resizebox{8cm}{8cm}{\includegraphics{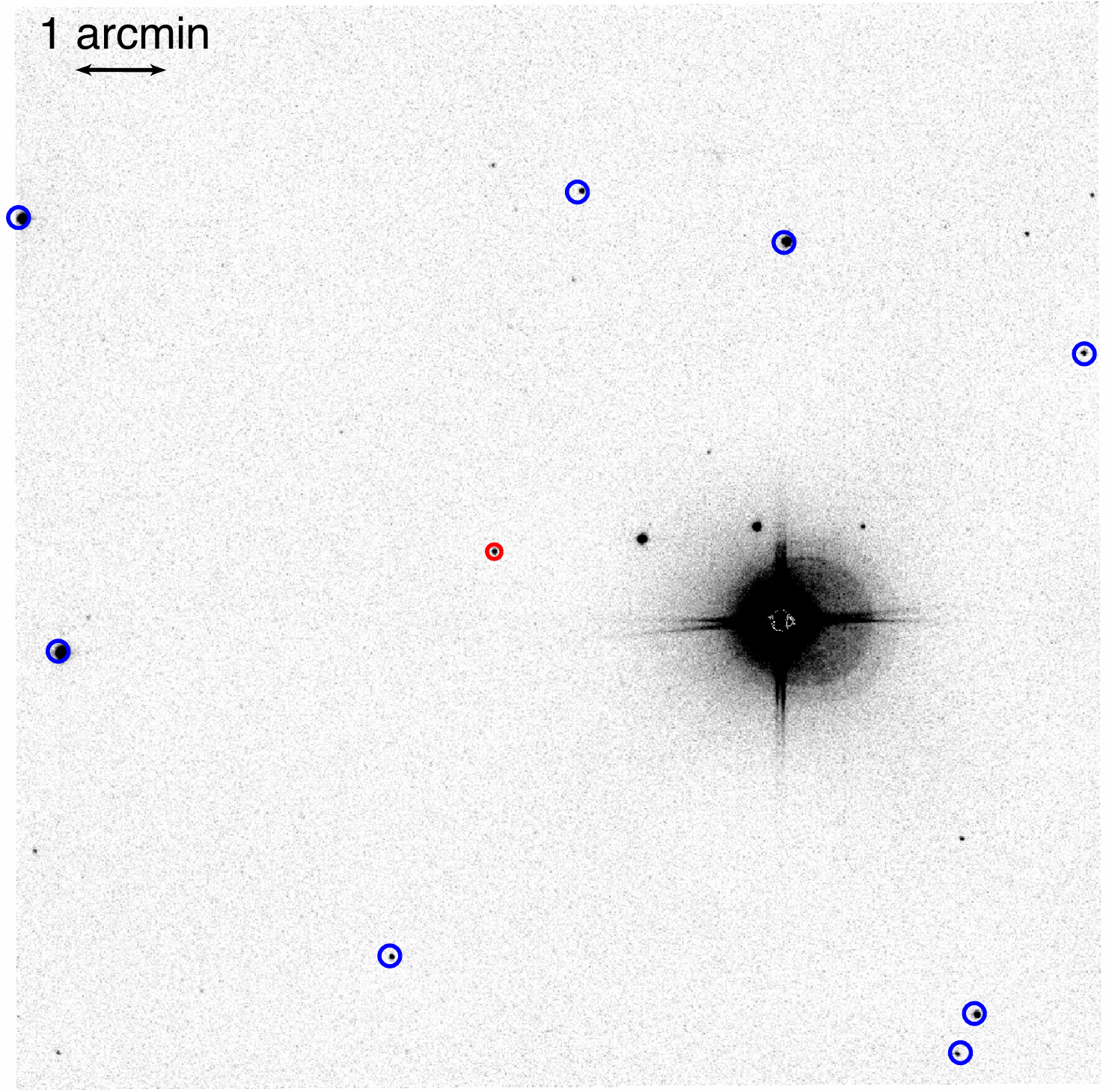}}
\resizebox{8cm}{8cm}{\includegraphics{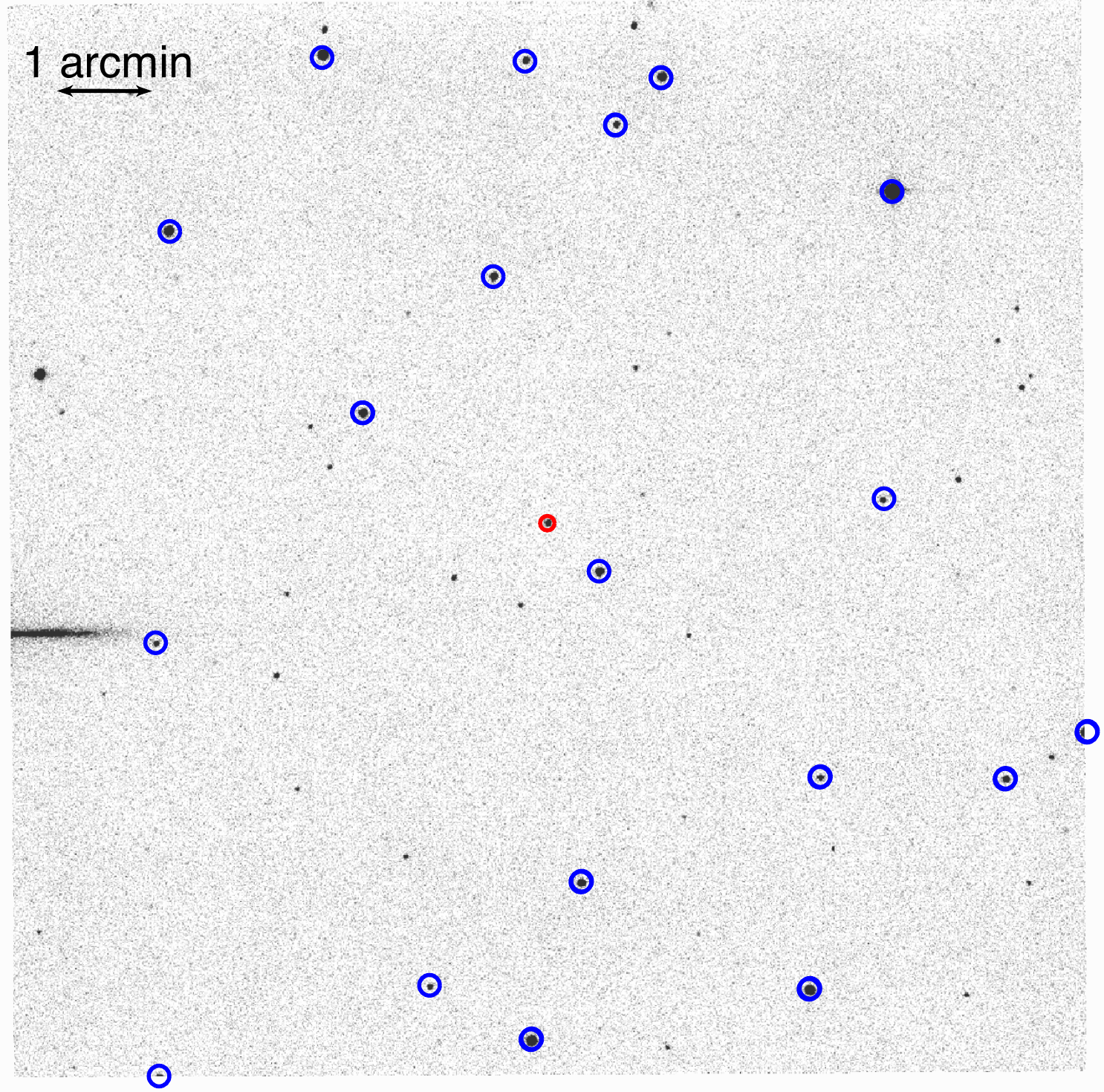}}
\caption{Example $B$-band images of 2MASS J1026 (left) and SDSS J1619 (right) obtained with the MDM 1.3m with an exposure time of 180s$\times3$ and 180s$\times2$, and seeing of 1.4$^{\prime \prime}$ and 1.5$^{\prime \prime}$, respectively. The target AGN and  comparison stars are marked with red and blue circles of diameter 7$^{\prime \prime}$ and 10$^{\prime \prime}$, respectively.}\label{Fig:field} 
\end{figure*}

\subsection{Photometric data}

Photometric observations were carried out 
at three different telescopes:

(I) MDM 1.3m, which is located at Kitt Peak, Tucson, Arizona, USA. We used the  1K$\times$1K Templeton CCD with a pixel scale 0.51$^{\prime \prime}$/pixel and the field of view (FOV) 8.7$^{\prime}$ $\times$ 8.7$^{\prime}$ (the 2K$\times$2K Echelle CCD, which trimmed with 1K$\times$1K pixels in the central region, was used on Feb. 23, 2017 and Nov. 6, 2018. The pixel scale is the same as that of the Templeton CCD). The median seeing during our monitoring campaign was 2.1$^{\prime \prime}$. We used the Johnson 4$^{\prime \prime}$ B filter\footnote{\url{http://mdm.kpno.noao.edu/instrumentation.html}} in both the MDM 1.3m and MDM 2.4m telescopes.

(II) Lemmonsan Optical Astronomy Observatory (LOAO) 1m, which is a robotic telescope located on Mt. Lemmon, Tucson, Arizona, USA. We used the 2K$\times$2K CCD with a 2 $\times$ 2 pixel on-chip binning, which provided a pixel scale 0.80$^{\prime \prime}$/pixel and the FOV 27$^{\prime}$ $\times$ 27$^{\prime}$. The median seeing was 3.2$^{\prime \prime}$. We used the $B$-band filter of ``Johnson/Cousins/Bessell Set''. 

(III) MDM 2.4m, which is located at Kitt Peak, Tucson, Arizona, USA. We used the MDM4K CCD (the R4K CCD, which has the same pixel size and pixel scale as those of the MDM4K, was used on Mar. 20$-$Apr. 8 and Dec. 8-9, 2017) with a pixel scale 0.273$^{\prime \prime}$/pixel and a FOV of 20$^{\prime}$ diameter circle (18.5$^{\prime}$ $\times$ 18.5$^{\prime}$ with the vignetting area). The median seeing during our monitoring campaign was 2.0$^{\prime \prime}$. 

In total 110 and 101 epochs of $B$-band photometric observations were carried out for 2MASS J1026 and SDSS J1619, respectively between November 2015 and July 2018 (see Figure \ref{Fig:field}). In addition, we used the $V$-band as a secondary filter in 27 epochs for 2MASS J1026 and in 22 epochs for SDSS J1619. 

We performed standard data reduction process using IRAF\footnote{\url{http://iraf.noao.edu/}} tasks, including bias subtraction, flat-fielding, and cosmic-ray removal using LA-cosmic\footnote{\url{http://www.astro.yale.edu/dokkum/lacosmic/}} \citep{2001PASP..113.1420V}.
For each image, we derived the astrometry solution, using Astrometry.net software \citep{2010AJ....139.1782L}.
After pre-processing, individual exposures were combined in the median, using {\scriptsize SWarp} \citep{2002ASPC..281..228B}. 
Then,
we measured the instrumental magnitudes of the target and comparison stars in the FOV with {\scriptsize SExtractor} \citep{1996A&AS..117..393B}.
Note that {\scriptsize SExtrator} calculates instrumental magnitudes in various aperture size (1$^{\prime \prime}$ $-$ 20$^{\prime \prime}$ as well as the `auto' aperture that covers $\ge$ 90\% of the total flux). While we used `auto' aperture for comparison stars in the FOV, for target AGNs, we additionally need to consider the effect of the host galaxy contribution. To determine the optimal aperture size for AGNs, we performed several tests 
by generating AGN light curves using various aperture sizes (1$^{\prime \prime}$ $-$ 20$^{\prime \prime}$ as well as the `auto' aperture).
In general, a smaller aperture excludes a larger amount of host galaxy light, while a larger aperture includes more host galaxy flux. Depending on the nightly varying seeing, the contribution of the host galaxy can also change, although the host galaxy contribution is relatively weak (see Section 4.2). While it is desirable to include the most of AGN flux using a large enough aperture, e.g., a factor of 3 larger than the seeing size, the error of the instrumental magnitude increases for a larger aperture size due to the increasing noise. 
We found that for the two AGNs, a large aperture (a factor of $\sim$3 larger than the seeing size) provides a consistent light curve, and we chose 7$^{\prime \prime}$ as an optimal aperture size for AGNs since it provides consistent and small measurement errors. We also visually inspected all the target images 
to check whether the fixed aperture is large enough to cover the target. 

We calculated the magnitude difference ($\Delta$mag) between instrumental magnitude and the expected magnitude of each comparison star, using the SDSS DR9 Catalog \citep{2012ApJS..203...21A}.  
The SDSS ugriz magnitudes were converted to the BVRI system
using the following formulas derived by Lupton (2005)\footnote{\url{https://www.sdss.org/dr15/algorithms/sdssUBVRITransform/\#Lupton2005}}:  
\begin{eqnarray} 
B = u - 0.8116*(u - g) + 0.1313 \\ 
V = g - 0.5784*(g - r) - 0.0038 
\end{eqnarray} 
We found that the calculated $B$-band magnitudes of the comparison stars are consistent with the $B$-band magnitude from APASS\footnote{\url{https://www.aavso.org/apass}} within an uncertainty of $\sim0.07$ magnitude.

The number of comparison stars used in the differential photometry of 2MASS J1026 (SDSS J1619) is 4 (11), 13 (37), and 66 (103), respectively for MDM 1.3m, MDM 2.4m, and LOAO images.  The mean of $\Delta$mag of these comparison stars provides the zero points while the standard deviation represents systematic uncertainty of the zero points. Finally, we rescaled the instrumental magnitude of AGN to the apparent magnitude using the calibrated zero point. The standard deviation of $\Delta$mag is added in quadrature to the instrumental magnitude error of AGN.  The details of the photometric measurements and calibration will be presented in a forthcoming paper (Cho, W. in preparation).


\subsection{Spectroscopic data}

Spectroscopic observations were carried out at two telescopes:

(I) Shane 3m telescope, which is located at the Lick observatory on Mt. Hamilton, California, USA. Lick observations were performed with a cadence of $\sim 20$ days using the Kast double spectrograph\footnote{\url{https://mthamilton.ucolick.org/techdocs/instruments/kast/}}. We used the red side of the spectroscope with a 600 lines/mm grating, covering the entire H$\beta$ emission line region. 
Prior to September 2016, we used the CCD with a 1,200$\times$400 pixel region, which provides a spatial scale of 0.78$^{\prime \prime}$/pixel and a spectral coverage of  4450 $-$ 7280 $\AA$ at 2.33 $\AA$/pixel. After that, a new 2K$\times$4K CCD with a spatial scale 0.43$^{\prime \prime}$/pixel and a spectral coverage of 4750 $-$ 8120 $\AA$ at 1.27 $\AA$/pixel has been used. We used a slit width of 4$^{\prime \prime}$ to minimize slit loss. The instrumental resolution is $R=624$, corresponding to a FWHM velocity resolution of 481 km s$^{-1}$. We obtained calibration frames, i.e., bias frames, dome flats and arc lamps (He, Ne, Ar and Hg-Cd) each night. The total exposure time per epoch was 30 min for 2MASS J1026 and 20 min for SDSS J1619, divided into 3 exposures. For observations at relatively low airmass $<1.3$, we used a fixed slit position angle (PA) while at high airmass ($>1.3$), AGNs were observed close to the parallactic angle.  

(II) MDM 2.4m; observations were carried out with a cadence of $\sim$1 month using two CCDs; MDM4K and R4K CCDs. A VPH blue grism with a spectral coverage of 3970 $-$ 6870 $\AA$ and a pixel scale of 0.715 $\AA$/pixel was used. We initially used a 3$^{\prime \prime}$ slit until Jan. 2017 since a 4$^{\prime \prime}$ slit was unavailable. To be consistent with Lick spectroscopy, we ordered a customized 4$^{\prime \prime}$ slit, which was used since Feb. 2017. The instrumental resolution is $R=617$ corresponding to a FWHM velocity resolution of 486 km s$^{-1}$. The effect of the resolution correction is insignificant on the measurement of broad emission lines (see section \ref{sec:line_width}). We obtained calibration frames, i.e., bias frames, dome flats, and Ar and Xe arc lamps each night. The total exposure time per epoch was 40 min and 60 min for 2MASS J1026 and SDSS J1619 respectively, divided into 2 exposures.

We performed standard spectroscopic data reduction using IRAF tasks, including overscan subtraction, bias and flat-fielding and cosmic-ray removal using LA-cosmic \citep{2001PASP..113.1420V}. 
One-dimensional spectra were extracted using the IRAF `apall' task. Considering the fact that the average seeing condition at Lick is worse than MDM, we used a 6$^{\prime \prime}$ aperture size for Lick spectra, in order to reduce the slit loss and the effect of nightly seeing variations, while for MDM spectra, we used a 4$^{\prime \prime}$ aperture. The wavelength solution was derived from a polynomial fit to the lamp spectra observed in each night and applied to the extracted spectra. For the MDM 2.4m observations, we applied an additional small linear shift to the wavelength scale to match the known wavelengths of strong sky absorption lines. Since we performed spectral decomposition, and fitted the H$\beta$ and [O III] emission lines of individual spectra as discussed in section \ref{sec:analysis}, our results are not affected by the small wavelength shift. We observed two spectrophotometric standard stars each night, and various standard stars were observed over the monitoring period. Flux calibration was initially performed using standard stars, however, the uncertainty of the flux calibration can be large due to the seeing effect, slit loss, and the systematic uncertainty in the flux calibration process. Thus, we rescaled each spectrum based on the flux of the narrow [O III] $\lambda$5007 emission line, which was assumed to be non-variable during the timescale of our monitoring campaign as described in the next section.

\begin{figure}
\centering
\resizebox{9cm}{6cm}{\includegraphics{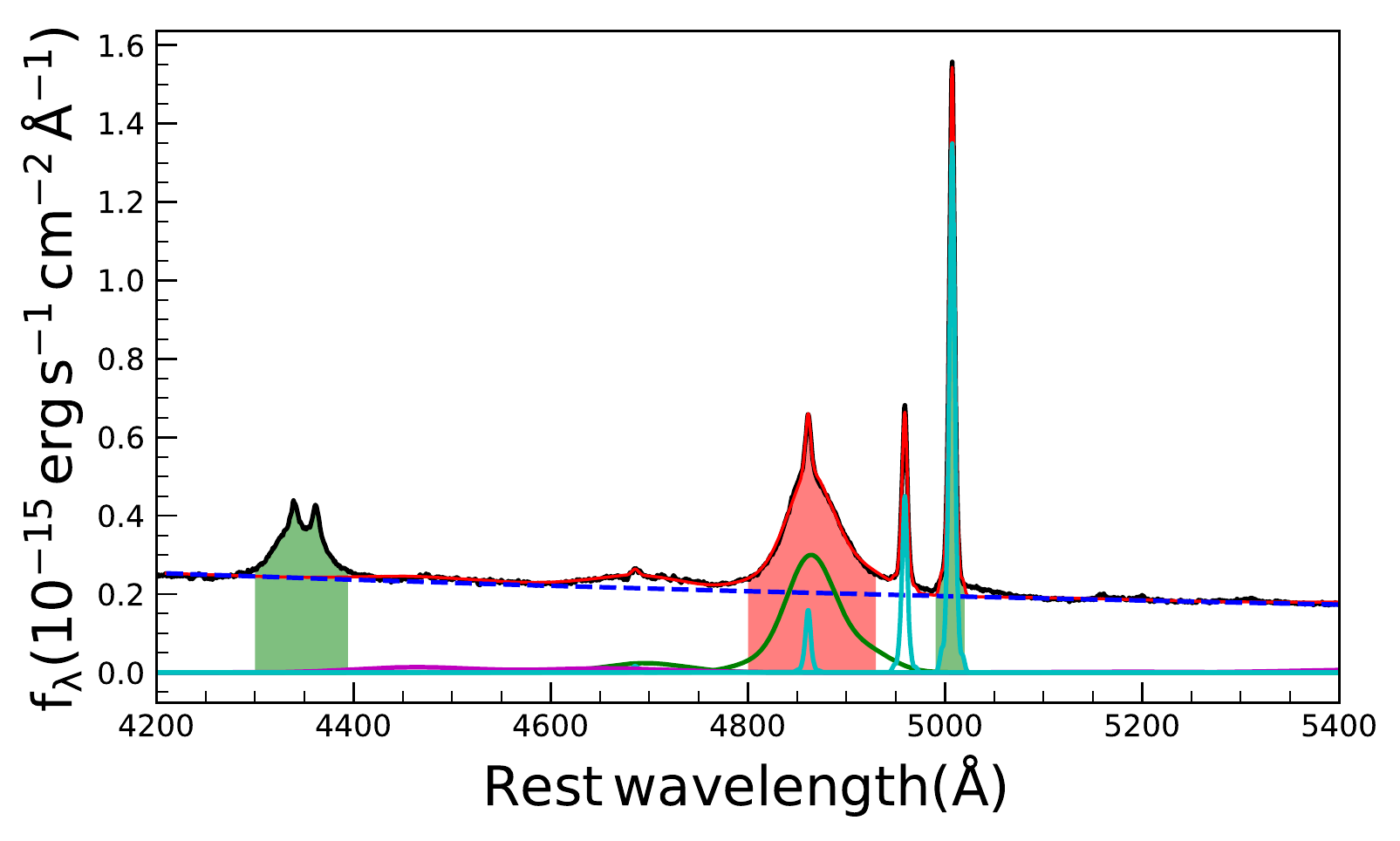}}
\resizebox{9cm}{6cm}{\includegraphics{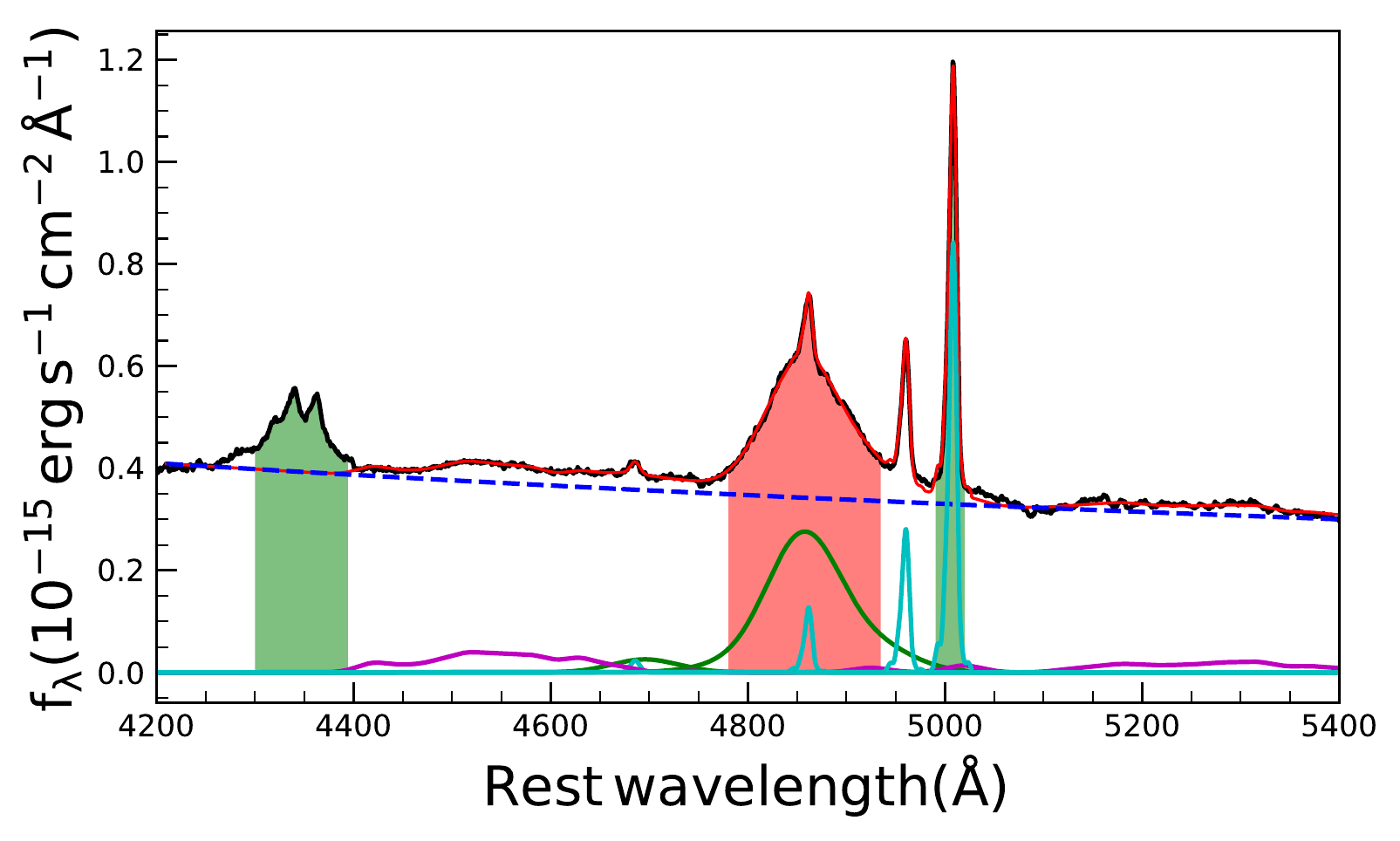}}
\caption{Spectral decomposition of the mean spectrum of 2MASS J1026 (top) and SDSS J1619 (bottom). The shaded regions represent the emission line integration windows.}\label{Fig:model_fit} 
\end{figure}

\begin{figure}
\centering
\resizebox{9cm}{12cm}{\includegraphics{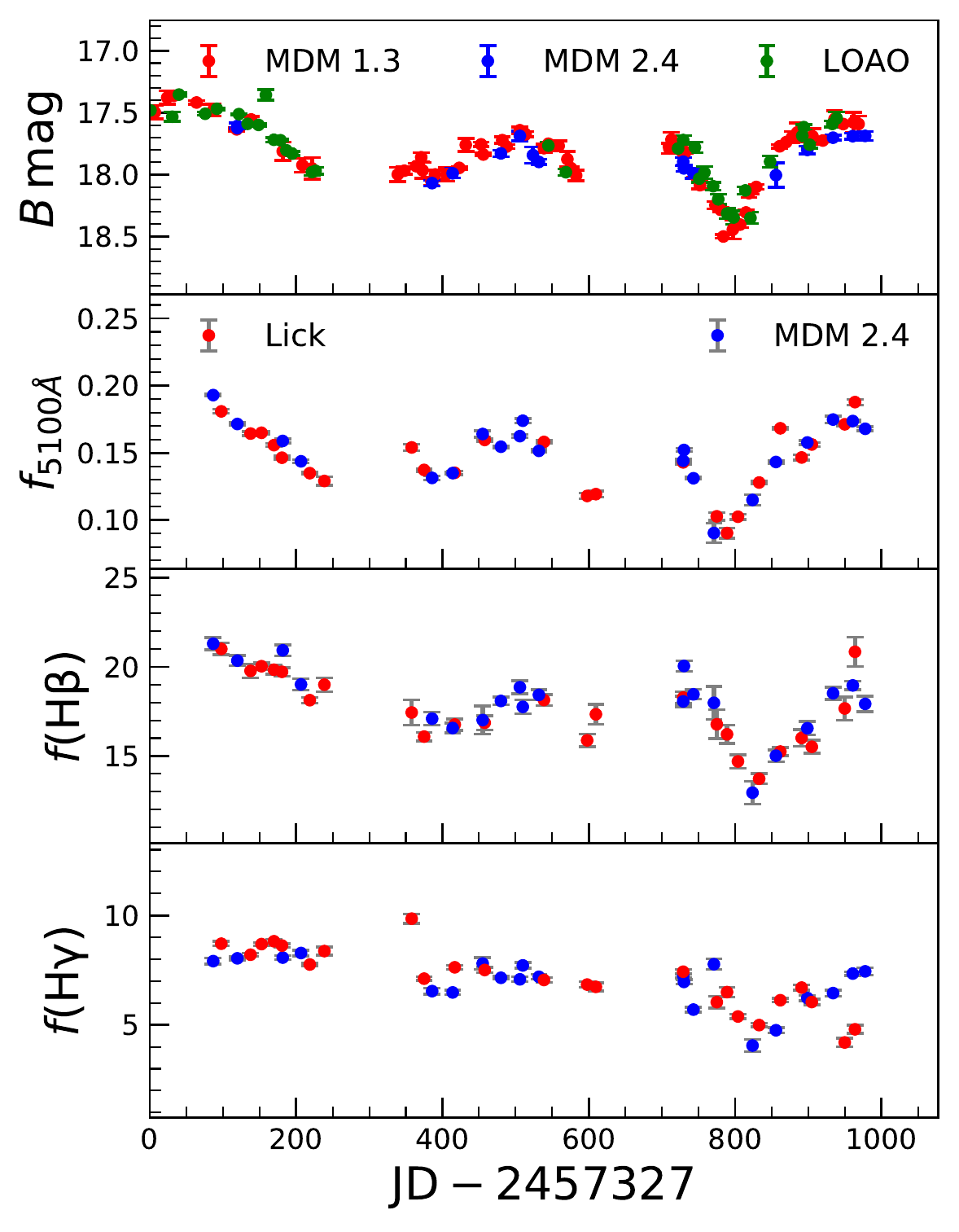}}
\caption{Light curves of 2MASS J1026. From top to bottom, $B$-band photometry, spectroscopic continuum at $5100$\AA, H$\beta$ and H$\gamma$ light curves are shown. The y-axis of spectroscopic continuum and the emission line light curves are in the units of $\mathrm{10^{-15}\, erg \, s^{-1}\, cm^{-2}\, \AA^{-1}}$ and $\mathrm{10^{-15}\, erg \, s^{-1}\, cm^{-2}}$, respectively.}\label{Fig:lc_ID17} 
\end{figure}

\begin{figure}
\centering
\resizebox{9cm}{12cm}{\includegraphics{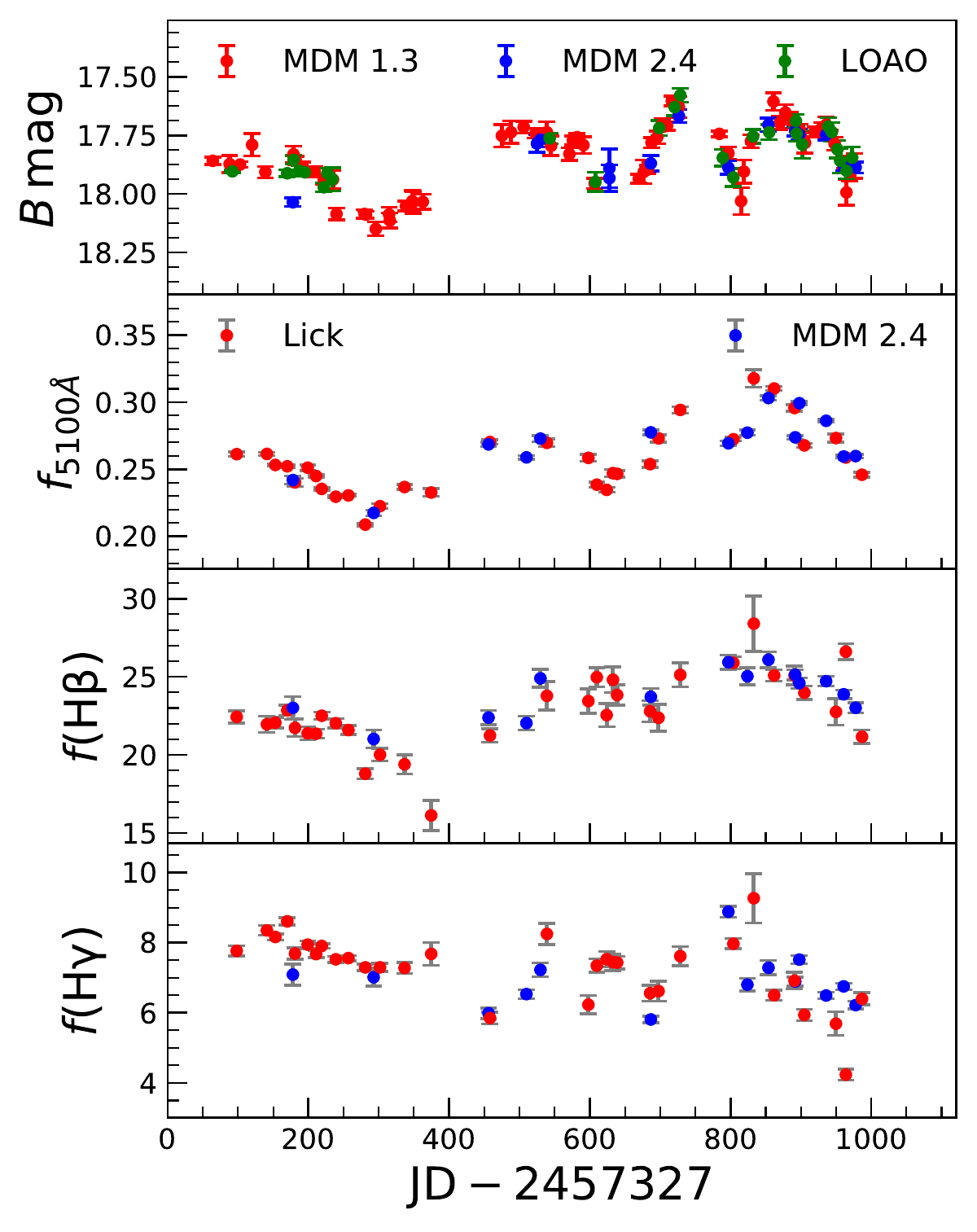}}
\caption{Light curves of SDSS J1619. Panels are same as in Figure \ref{Fig:lc_ID17}.}\label{Fig:lc_ID43} 
\end{figure}

\section{Results and analysis}\label{sec:analysis}

\subsection{Spectral decomposition}\label{sec:spectra_analysis}

We performed multi-component spectral fitting to measure the continuum and emission line properties using individual spectra \citep[see][]{2012ApJ...747...30P,2017ApJ...839...93P,2017ApJ...847..125P}. Focusing on the H$\beta$ line region, we first modeled the continuum by combining a power-law component representing AGN continuum and a model of Fe II emission blends. For Fe II, we used the template from \citet{2010ApJS..189...15K}, which provided a better fitting of diverse Fe II emission blends compared to other available Fe II templates \citep[see][]{2017ApJ...839...93P}. Since stellar absorption lines were very weak in the observed spectra, we did not model the stellar contribution. We performed the nonlinear Levenberg-Marquardt least-squares minimization using the IDL code {\scriptsize MPFIT} \citep{2009ASPC..411..251M}. The best power-law component and Fe II models were then subtracted, leaving emission lines. 

In the H$\beta$ region, we modeled [O III] $\lambda 5007$ in the spectral range of $4979-5022 \AA$ using a 10th order Gauss-Hermite polynomial \citep{1994MNRAS.270..271V}. The [O III] $\lambda 4969$ line was fitted using the same velocity profile and fixing the flux ratio of [O III] $\lambda4959$ and [O III] $\lambda5007$ to its theoretical value. The narrow H$\beta$ component was modeled using the [O III] profile with the flux as a free parameter. The broad H$\beta$ component was modeled in the region of $4791-4931 \AA$ using a 4th order Gauss-Hermite polynomial. In the case of the He II $\lambda4686$ line, we used two Gaussian components; one for the broad and another for the narrow components. 
Since it was difficult to properly decompose the broad and narrow H$\beta$ components in each epoch, we used the total H$\beta$ line flux in order to minimize any systematic uncertainty caused by the decomposition of the two components. The H$\gamma$ line region includes fluxes from the broad and narrow H$\gamma$ and [O III] $\lambda4363$. However, fitting individual components are difficult without very high S/N spectra. Hence, instead of modeling the individual lines, we measured the emission line flux by subtracting the best-fit continuum (AGN power-law and Fe II) and integrating the emission flux directly from the continuum-subtracted spectra. 

Emission line wings were not well constrained in some epochs due to low S/N. To minimize the systematic uncertainty, we integrated the line flux within a limited window to avoid the line wings, instead of using the entire line profiles of [O III], H$\beta$ and H$\gamma$. The integration window of each emission line is defined in Table \ref{Table:wavelength_range} (see Figure \ref{Fig:model_fit}). We rescaled all spectra by forcing [O III] $\lambda 5007$ flux to be constant, assuming the [O III] flux did not vary during our campaign. We used high-quality spectra taken in the best weather condition to estimate the [O III] $\lambda 5007$ flux for normalization. However, the choice of the reference [O III] $\lambda 5007$ flux has no effect on the final results since the continuum luminosity was measured from the mean spectra after re-calibration with the photometry results (see section \ref{sec:Size-luminosity}).

 \begin{table}
 \caption{Integration window of each emission line.}
	\begin{center}
	\hspace*{-1.1cm}
 	\resizebox{1.1\linewidth}{!}{%
     \begin{tabular}{ l l l}\hline \hline 
     Object                     & emission line  &  rest-frame wavelength range (\AA)  \\\hline
     2MASS J1026    & H$\beta$     & $4800-4930$ \\
                                & H$\gamma$    & $4300-4395$  \\
                                & [O III]      & $4990-5020$  \\
     SDSS J1619   & H$\beta$     & $4780-4935$ \\
                                & H$\gamma$    & $4300-4395$ \\ 
                                & [O III]      & $4990-5020$  \\\hline            
        \end{tabular} } 
        \label{Table:wavelength_range}
        \end{center}
    \end{table}

 \begin{table}
 \caption{Photometric data. Columns are (1) object name, (2) Julian date, (3) $B$-band magnitude and (4) telescope name. The table is available in its entirety in a machine-readable form in the online journal. A portion is shown here for guidance regarding its form and content.}
	\begin{center}
	\hspace*{-1.2cm}
 	\resizebox{1.0\linewidth}{!}{%
     \begin{tabular}{ l l l l}\hline \hline 
     Object                     & JD           &  magnitude  & Telescope \\
     (1)                        & (2)          & (3) 	     & \\\hline
    2MASS J1026	             	& 2457327.9703 & 17.48 $\pm$ 0.02 & LOAO \\
    							& 2457333.9734 & 17.49 $\pm$ 0.05 & MDM13 \\
    							& 2457350.9654 & 17.37 $\pm$ 0.05 & MDM13 \\
            
     \hline
        \end{tabular} } 
        \label{Table:lc_phot}
        \end{center}
    \end{table}
    
     \begin{table}
     \caption{Spectroscopic data. Columns are (1) object name, (2) Julian date (3) monochromatic flux at 5100 \AA \, in the units of $\mathrm{10^{-15}\, erg \, s^{-1}\, cm^{-2}\, \AA^{-1}}$, (4) and (5) are the H$\beta$ and H$\gamma$ line flux, respectively in the units of $\mathrm{10^{-15}\, erg \, s^{-1}\, cm^{-2}}$ and (6) telescope name. The table is available in its entirety in a machine-readable form in the online journal. A portion is shown here for guidance regarding its form and content.}
    	\begin{center}
    	\hspace*{-1.2cm}
     	\resizebox{1.1\linewidth}{!}{%
         \begin{tabular}{ l l l l l l}\hline \hline 
         Object                  & JD                &  $f_{5100}$		& $f ({\mathrm{H\beta}})$  & $f ({\mathrm{H\gamma}})$ & telescope  \\
         (1)                     & (2)               & (3)              & (4)           & (5)  & (6) \\\hline
         2MASS J1026             & 2457413.7587 & 0.193 $\pm$ 0.001 & 21.300 $\pm$ 0.349 & 7.914 $\pm$ 0.138 & MDM24 \\
         						 & 2457424.8030 & 0.181 $\pm$ 0.002 & 21.008 $\pm$ 0.332 & 8.709 $\pm$ 0.101 & Lick \\
         						 & 2457446.7722 & 0.171 $\pm$ 0.001 & 20.349 $\pm$ 0.295 & 8.041 $\pm$ 0.074 & MDM24 \\
         
         \hline 
            \end{tabular} } 
            \label{Table:lc_spec}
            \end{center}
        \end{table}

 \begin{table}
 \caption{Variability information. Columns are (1) object name, (2) light curve, (3) rms variability amplitude, (4) maximum to minimum flux variation, (5) number of epochs, (6) mean sampling rate (days) over the entire campaign, and (7) typical uncertainty in the light curve.}
	\begin{center}
	\hspace*{-1.2cm}
 	\resizebox{0.9\linewidth}{!}{%
     \begin{tabular}{ l l l l l l l }\hline \hline 
     Object                     & light curve  &  $F_{\mathrm{var}}$ & $R_{\mathrm{max}}$ & epochs & sampling & uncertainty (\%)     \\
     (1)            & (2)        &  (3)      & (4)             & (5) & (6) & (7)\\\hline
    2MASS J1026		& $B$-band   &   0.21    & $2.87 \pm 0.05$ & 110 & 9   & 2.8 (0.03 mag) \\
    			    & $f_{5100}$ &   0.16    & $2.13 \pm 0.08$ & 45  & 20  & 1.1  \\
     				& H$\beta$   &   0.11    & $1.64 \pm 0.08$ & $-$ & $-$ & 2.2\\
                    & H$\gamma$  &   0.17    & $2.42 \pm 0.17$ & $-$ & $-$ & 1.9\\
     SDSS J1619     & $B$-band   &   0.11    & $1.69 \pm 0.06$ & 101 & 9   & 2.8 (0.03 mag) \\
     		        & $f_{5100}$ &   0.09    & $1.52 \pm 0.03$ & 46  & 19  & 0.7\\
     				& H$\beta$   &   0.09    & $1.76 \pm 0.15$ & $-$ & $-$ & 2.3\\
                    & H$\gamma$  &   0.12    & $2.18 \pm 0.18$ & $-$ & $-$ & 2.5\\           
     \hline 
        \end{tabular} } 
        \label{Table:var_info}
        \end{center}
    \end{table}
    
\subsection{Variability}    

The final light curves are shown in Figure \ref{Fig:lc_ID17} for 2MASS J1026 and Figure \ref{Fig:lc_ID43} for SDSS J1619. The photometric and spectroscopic measurements are given in Tables \ref{Table:lc_phot} and \ref{Table:lc_spec}, respectively. Note that the presented spectroscopic light curves were 
before re-calibration with the photometry (see section \ref{sec:Size-luminosity}). 
We calculated the fractional root-mean-square (rms) variability amplitude $F_{\mathrm{var}}$ \citep{2002ApJ...568..610E} using 
\begin{equation}
F_{\mathrm{var}} = \frac{\sqrt{\sigma^2 - <\delta^2>}}{<f>},
\end{equation}
where $\sigma^2$ is the variance, $<\delta^2>$ is the mean square error, and $<f>$ is the arithmetic mean of the light curves. The values of $F_{\mathrm{var}}$ for both objects are about $0.1$ (see Table \ref{Table:var_info}), indicating significant variability in the continuum as well as H$\beta$ and H$\gamma$ lines. The variation in H$\gamma$ is larger than in H$\beta$ as expected from the photo-ionization calculation \citep{2004ApJ...606..749K} and past reverberation mapping studies of other AGNs \citep[e.g.,][]{2010ApJ...716..993B}. Moreover, the $R_{\mathrm{max}}$, which is the ratio of maximum to minimum flux variation is about 64\% and 76\% in H$\beta$, respectively for 2MASS J1026  and SDSS J1619. In the case of H$\gamma$, the line flux increases more than a factor of 2 in both objects. 

\begin{figure}
\centering
\resizebox{7.0cm}{6cm}{\includegraphics{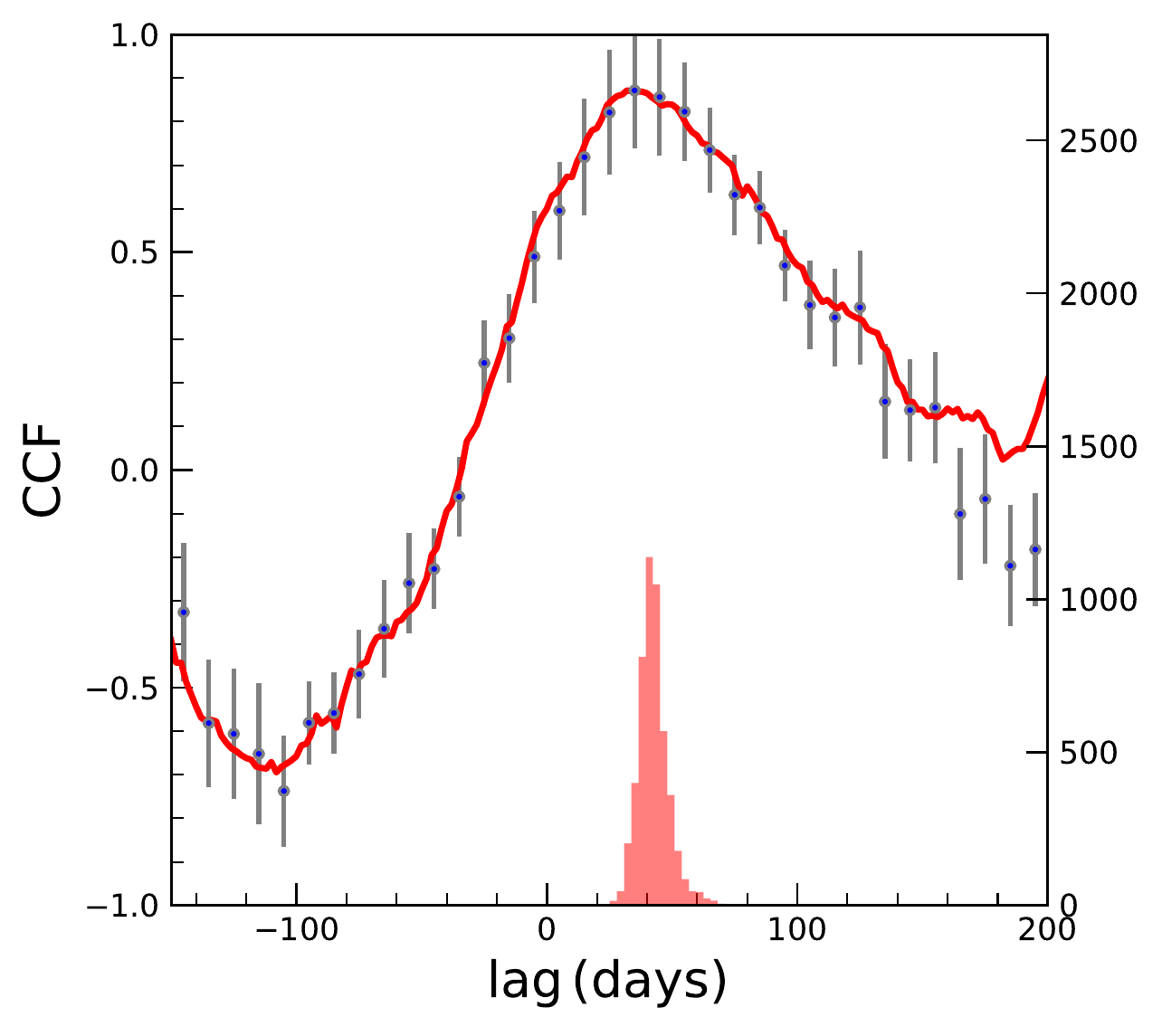}}
\resizebox{7.0cm}{6cm}{\includegraphics{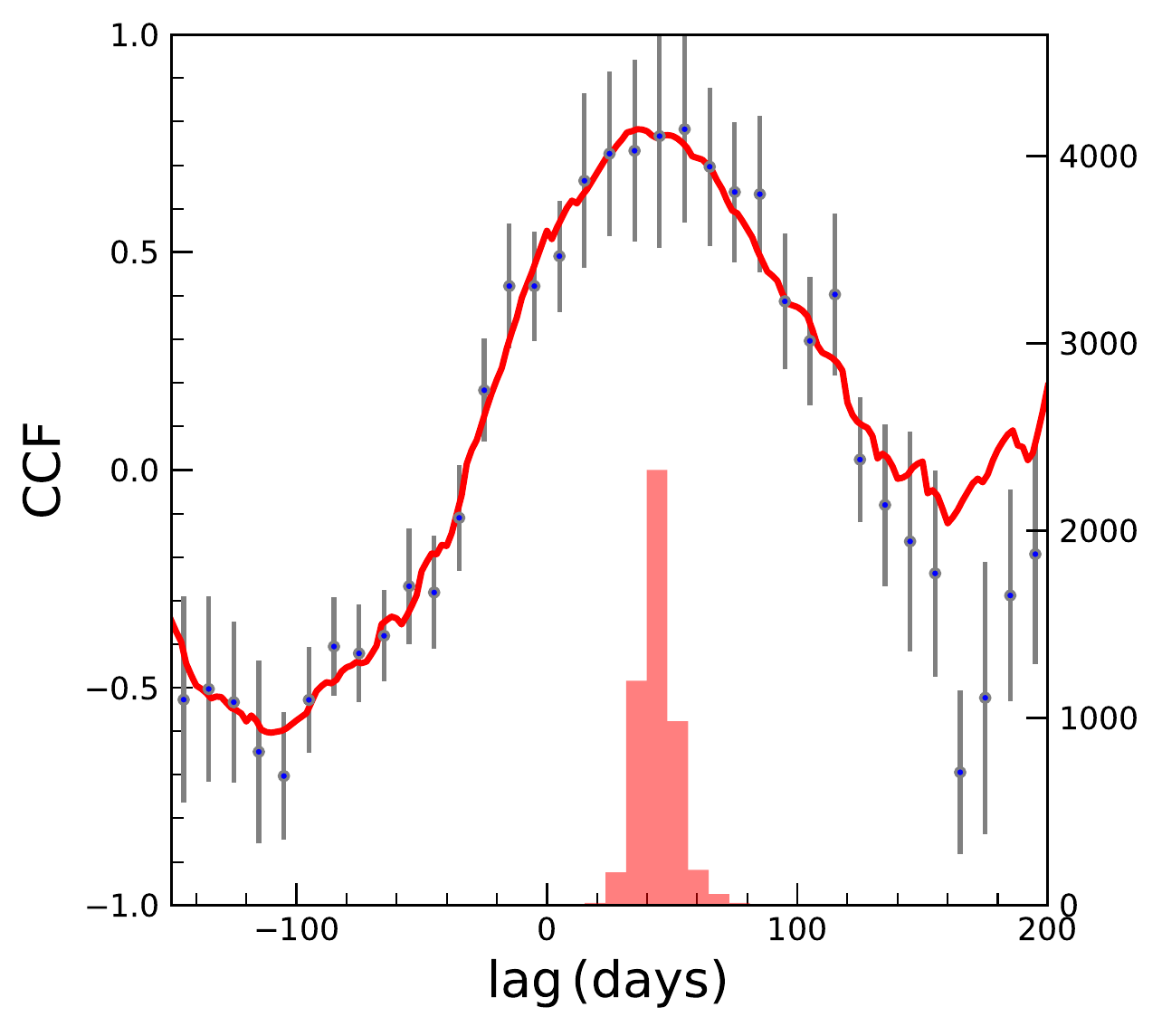}}
\resizebox{7.0cm}{6cm}{\includegraphics{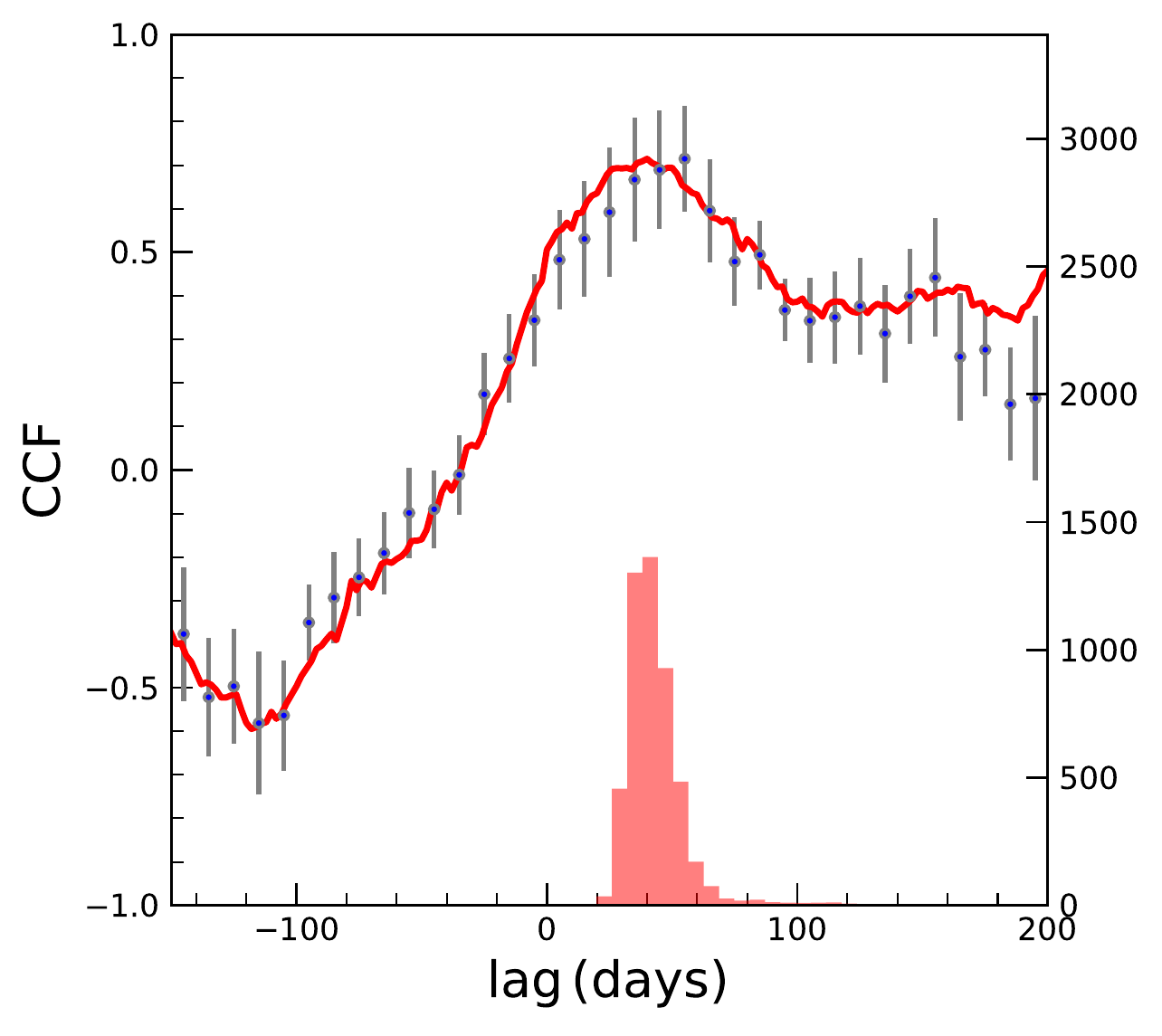}}
\caption{Cross-correlation of $B$ vs H$\beta$ (top), $f_{5100}$ vs H$\beta$ (middle), and $B$ vs H$\gamma$ (bottom) for 2MASS J1026. The ICCF is shown by the solid line while the points denote the DCF. The centroid distribution obtained from the ICCF is shown by the histogram. The last two Lick data points were not considered for $B$ vs H$\gamma$ correlation.}\label{Fig:ID17_ccf} 
\end{figure}

\begin{figure*}
\centering
\resizebox{7cm}{6cm}{\includegraphics{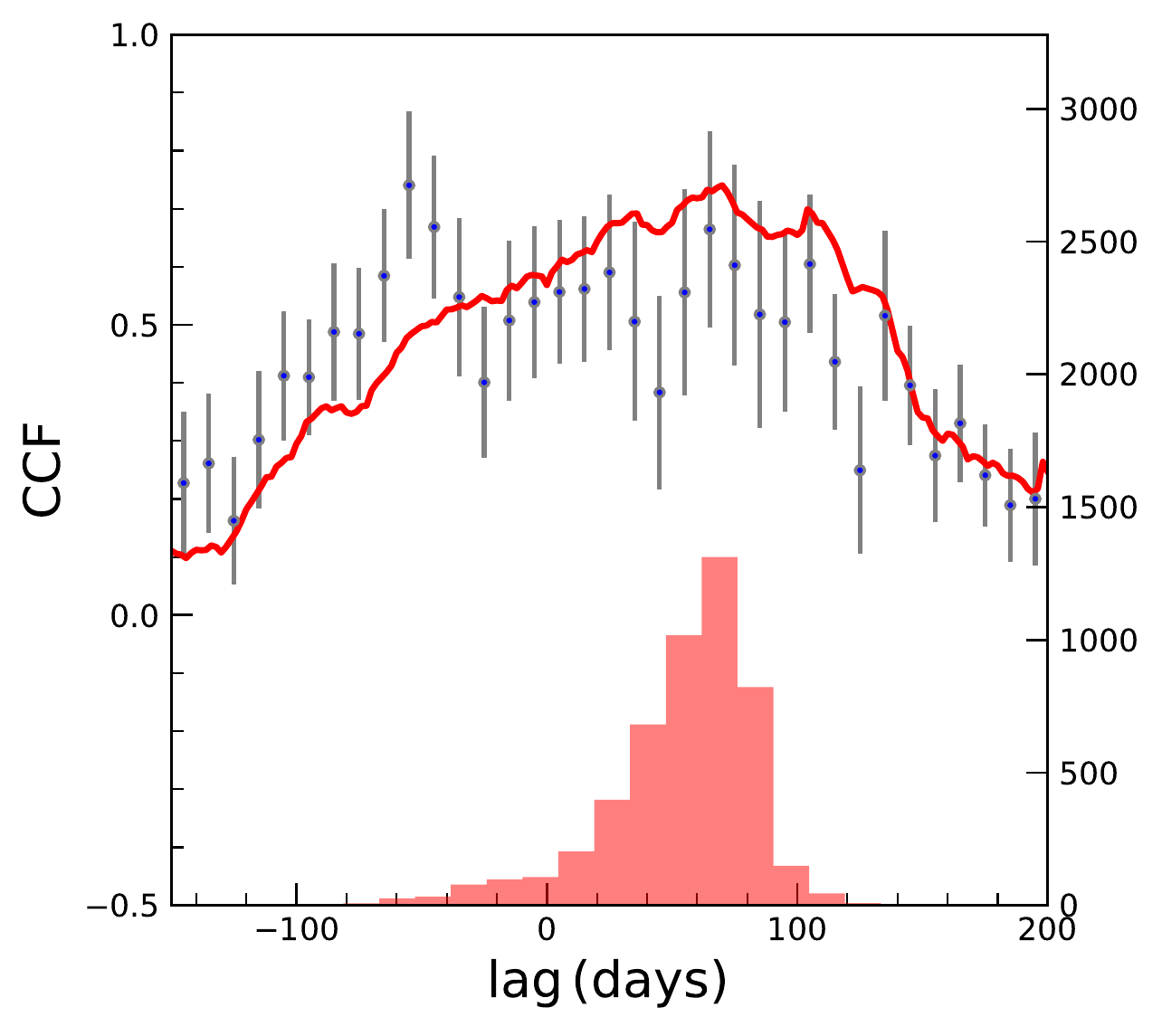}}
\resizebox{7cm}{6cm}{\includegraphics{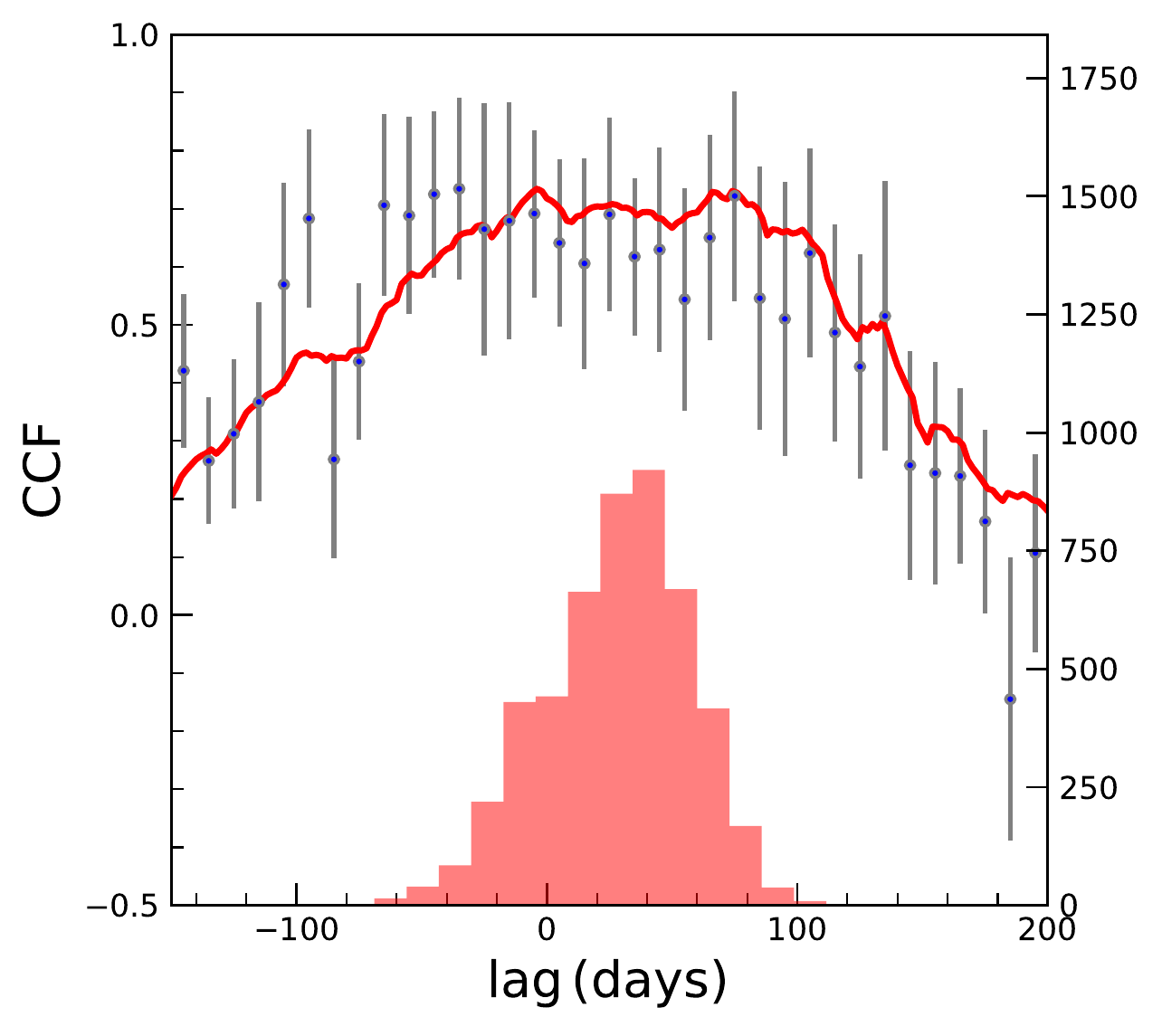}}
\resizebox{7cm}{6cm}{\includegraphics{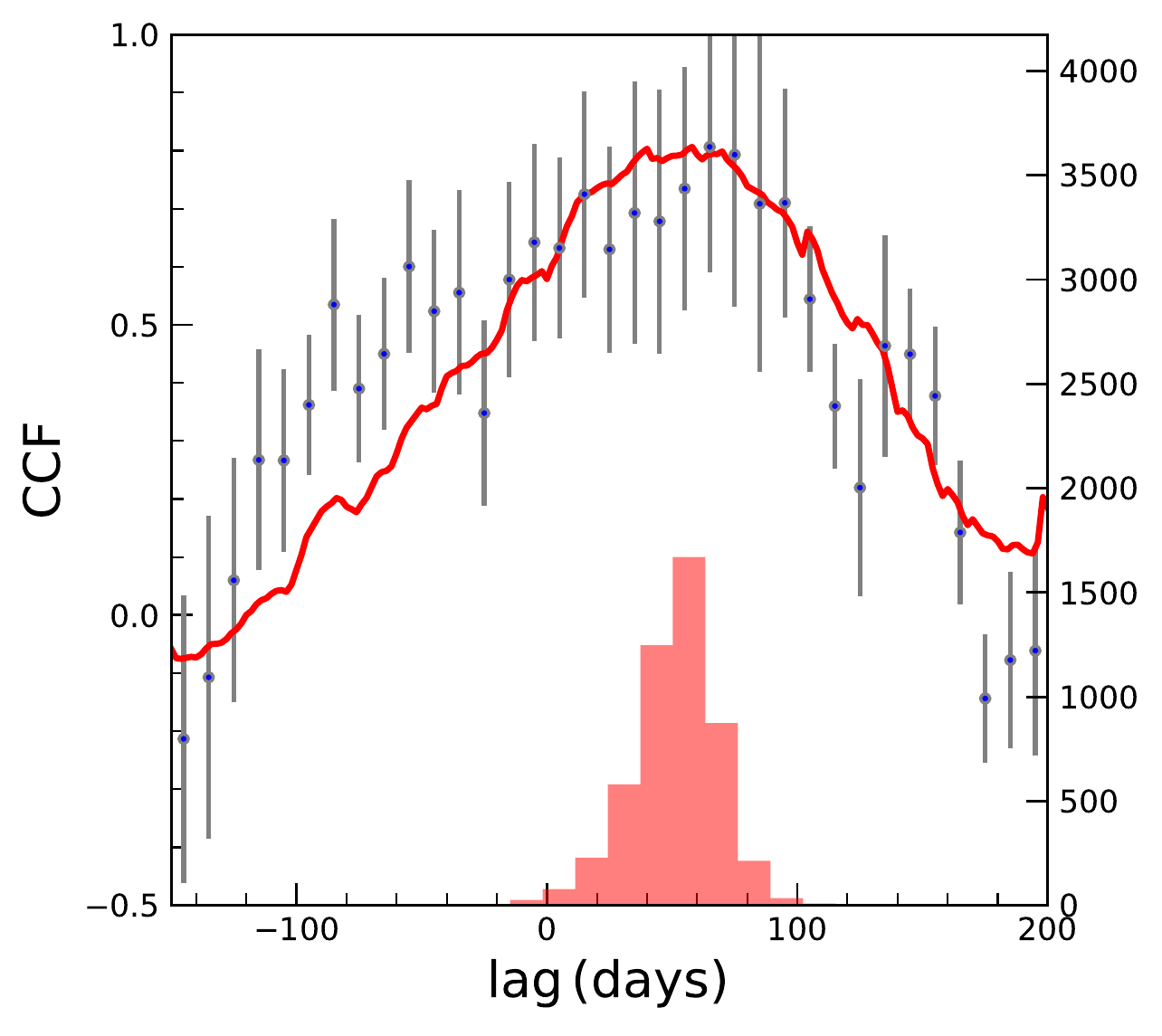}}
\resizebox{7cm}{6cm}{\includegraphics{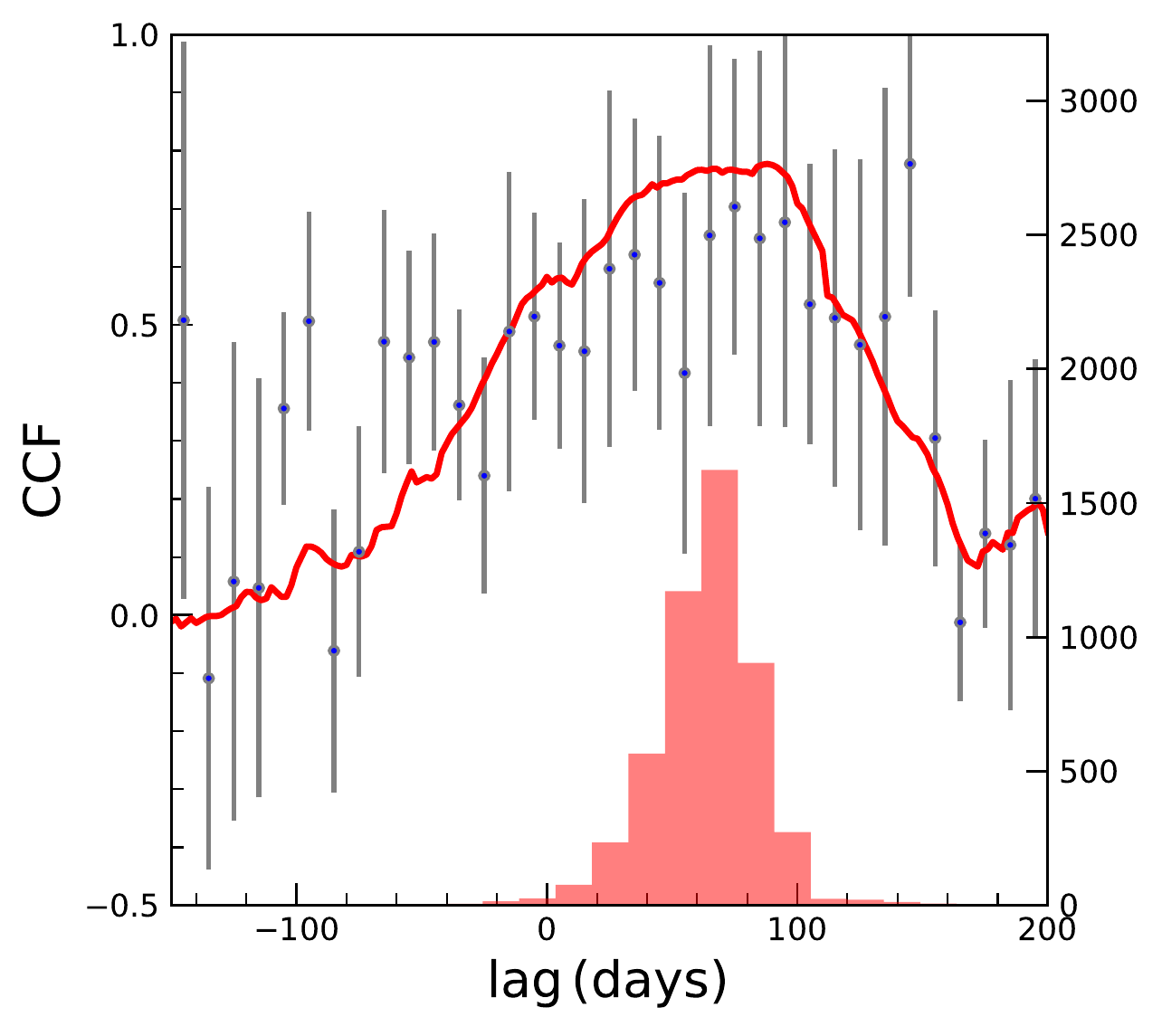}}
\caption{Left: Cross-correlation of $B$ vs H$\beta$ for SDSS J1619, using the entire $B$ and H$\beta$ light curves (top) and a part of the $B$ and H$\beta$ light curves until JD=2457327+800 (bottom). Right: same as the left panel but for $f_{5100}$ vs H$\beta$. The ICCF is shown by the solid line while the points denote the DCF. The centroid distribution obtained from the ICCF is shown by the histogram.}\label{Fig:ID43_ccf} 
\end{figure*}

\subsection{Time lag}\label{sec:lag}

To measure the time delay, we used the cross-correlation technique \citep{1987ApJS...65....1G,1994PASP..106..879W,2004ApJ...613..682P}, as commonly used in the reverberation mapping analysis. We performed the interpolated cross-correlation analysis using python code {\scriptsize pyCCF} \citep{2018ascl.soft05032S}, which is based on the methodology of \citet{2004ApJ...613..682P}. First, we interpolated the continuum light curve and calculated the cross-correlation function (CCF) between the interpolated continuum and the emission line light curves. Second, we re-calculated the CCF between the observed continuum light curve and interpolated emission line light curve. The average of the two CCFs provided the final interpolated cross-correlation function (ICCF). In addition, we also measured the discrete correlation function \citep[DCF;][]{1988ApJ...333..646E}. We calculated the centroid of the CCF ($\tau_{\mathrm{cent}}$) using the points with CCF $>$ 0.8 $\times$ peak of the CCF. The ICCF (solid) and DCF (blue points) are presented in Figure \ref{Fig:ID17_ccf} for 2MASS J1026 and in Figure \ref{Fig:ID43_ccf} for SDSS J1619. 

Using the flux randomization and random subset sampling (FR/RSS) method \citep{1987ApJS...65....1G,1994PASP..106..879W,2004ApJ...613..682P}, we performed Monte Carlo realizations of the light curves to estimate the uncertainty in the lag measurement. In each realization, we randomly selected the same number of points in the light curve, and if one epoch is selected n times, the uncertainty of the flux was reduced by n$^{1/2}$. We created 5000 mock light curves by adding Gaussian noise based on the flux uncertainty at each epoch, and performed the cross-correlation and estimated lag for each mock light curve. We took the median of the resulting distribution as an estimation of the final $\tau_{\mathrm{cent}}$, and the uncertainty on each side is measured at 1$\sigma$ (68\%) from the median. The distribution of $\tau_{\mathrm{cent}}$ obtained by the ICCF method is shown in Figures \ref{Fig:ID17_ccf} and \ref{Fig:ID43_ccf} (see Table \ref{Table:ccf}). Both ICCF and DCF methods provide consistent lag within the uncertainty. Since ICCF has been extensively used in previous reverberation mapping studies, we used the lag based on the ICCF method for further analysis.

For 2MASS J1026, we obtained the $B$-band to H$\beta$ lag as $41.8^{+4.9}_{-6.0}$ days in the observed-frame, and the $B$-band to H$\gamma$ lag as $41.1^{+7.1}_{-9.9}$ days based on the ICCF method. If we instead used the continuum $f_{5100}$ light curve from spectroscopy, we obtained 
a consistent H$\beta$ lag as $43.4^{+6.4}_{-7.5}$ days. 

In the case of SDSS J1619, we obtained a H$\beta$ lag of $60.1^{+33.1}_{-19.0}$ days. However, there is a discrepancy that the $B$-band light curve shows a dip around $800-850$ days, while the 5100\AA\ light curve shows a much weaker dip. The dip is not present in the H$\beta$ light curve. Due to this discrepancy, the CCF shows a broad peak and the lag is not well constrained. 
Thus, we decided to use a part of the light curves by excluding the epochs obtained after JD$= 2457327+800$, and recalculated the lag. In this case, we obtained a lag of $52.6^{+17.6}_{-14.7}$ days. If we use the $f_{5100}$ light curve instead, the lag is not well constrained mainly due to the worse time sampling and low variability. Since the CCF obtained with the limited light curves provides a well-defined peak and a better constrained lag, 
we adopt this measurement as the final result. In the case of H$\gamma$, the CCF is relatively flat, preventing us from obtaining a reliable lag measurement. Since the sampling of the photometric light curves is twice better than that of the spectroscopic (i.e., 5100\AA) light curves,
we adopted the lag measurements with the $B$-band light curves as the best values.

 \begin{table}
 \caption{Cross-correlation analysis results. Columns are as follows (1) object name, (2) light curves for cross-correlation analysis, lags based on ICCF (3), and DCF (4) centroid distributions. All the lags are in the observed-frame.}
	\begin{center}
	\hspace*{-1.2cm}
 	\resizebox{1.1\linewidth}{!}{%
     \begin{tabular}{ l l l l}\hline \hline 
     Object         & light curves       &  $\tau_{\mathrm{cent}}$ (ICCF)    & $\tau_{\mathrm{cent}}$ (DCF) \\
                    &                    &    (days)                         & (days) \\ 
      (1)           &  (2)               &    (3)                            &  (4)    \\ \hline
    2MASS J1026		& $B$ vs. H$\beta$         & $41.8^{+4.9}_{-6.0}$     & $42.5^{+12.0}_{-6.5}$  \\
    				& $f_{5100}$ vs. H$\beta$  & $43.4^{+6.4}_{-7.5}$     & $50.1^{+15.8}_{-3.4}$ \\
     			    & $B$ vs. H$\gamma$        & $41.1^{+7.1}_{-9.9}$     & $54.9^{+25.8}_{-0.5}$ \\
    SDSS J1619      & $B$ vs. H$\beta$ (full)  & $60.1^{+33.1}_{-19.0}$   & $41.1^{+24.8}_{-45.3}$\\
                    & $f_{5100}$ vs. H$\beta$ (full) & $30.7^{+35.2}_{-24.7}$ & $--$ \\
    				& $B$ vs. H$\beta$ (part)  & $52.6^{+17.6}_{-14.7}$   & $57.9^{+23.7}_{-17.8}$\\
    				& $f_{5100}$ vs. H$\beta$ (part) & $65.0^{+20.2}_{-16.9}$ & $--$ \\
              
     \hline 
        \end{tabular} } 
        \label{Table:ccf}
        \end{center}
    \end{table}

\begin{figure}
\centering
\resizebox{9cm}{7cm}{\includegraphics{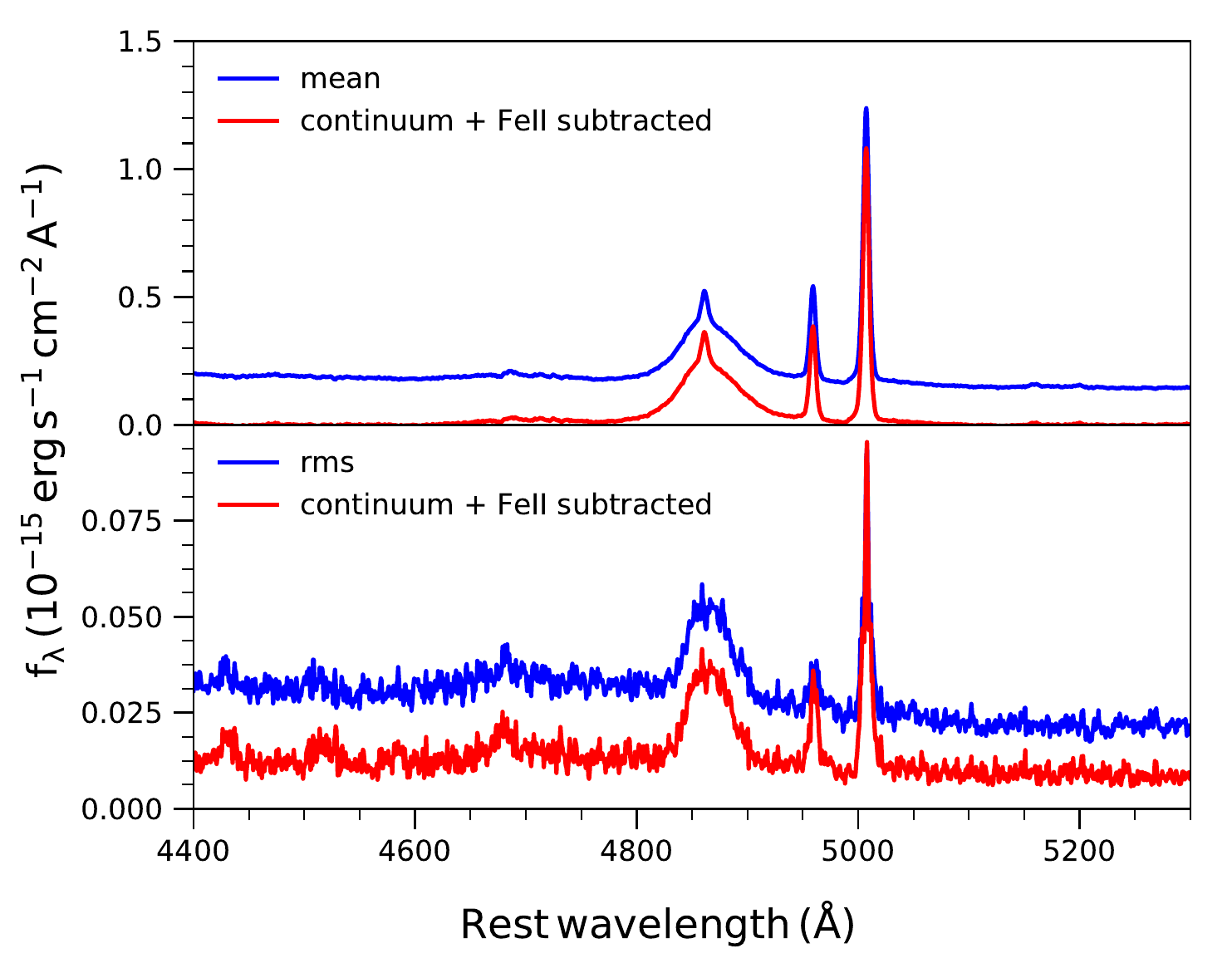}}
\resizebox{9cm}{7cm}{\includegraphics{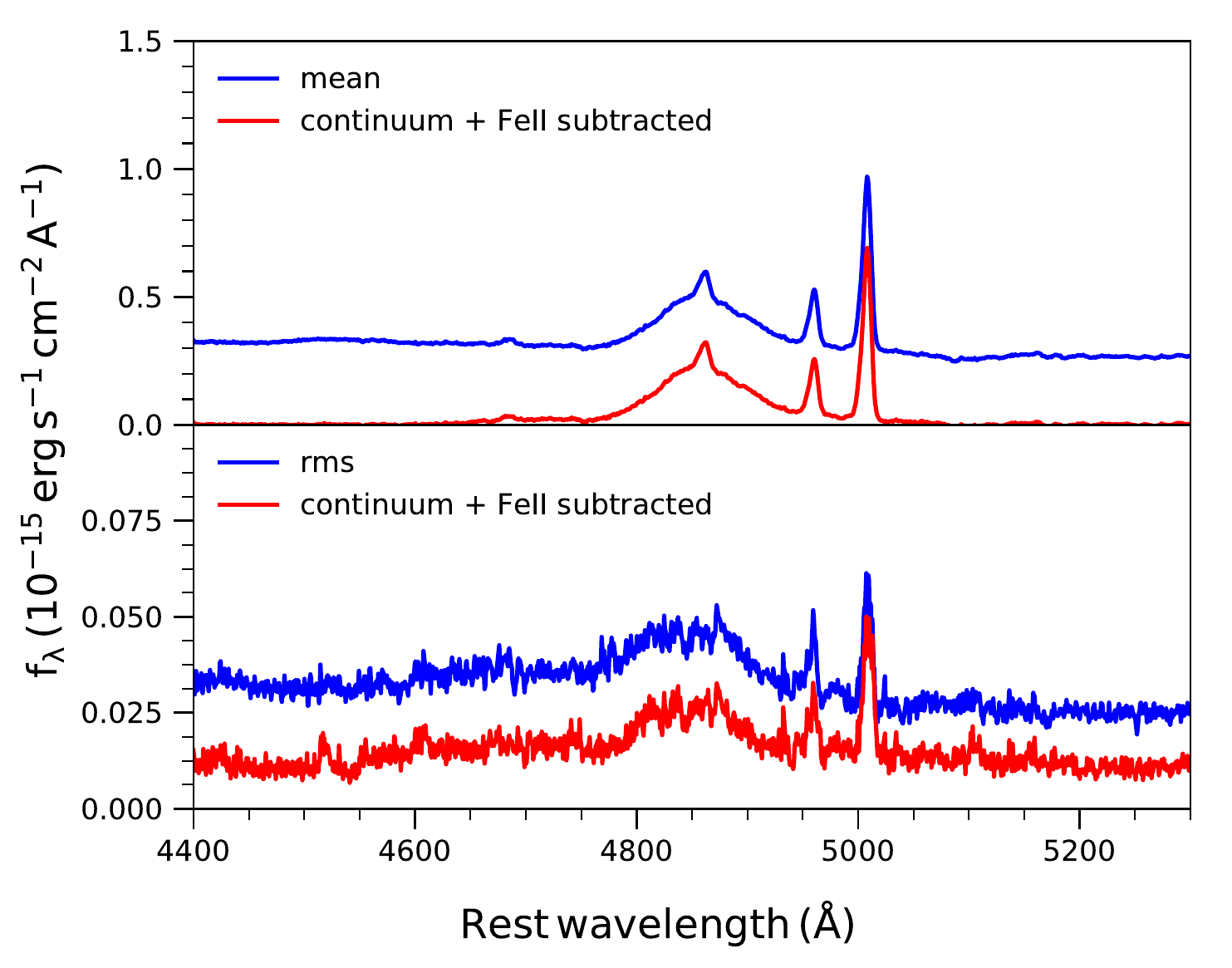}}
\caption{Mean and rms spectra of 2MASS J1026 (top) and SDSS J1619 (bottom). The mean spectrum calculated from the rescaled individual spectra (blue), and the mean spectrum after removing continuum and Fe II contribution from individual spectra (red) are shown in each panel.}\label{Fig:mean_rms} 
\end{figure}

\begin{figure}
\centering
\resizebox{9cm}{7cm}{\includegraphics{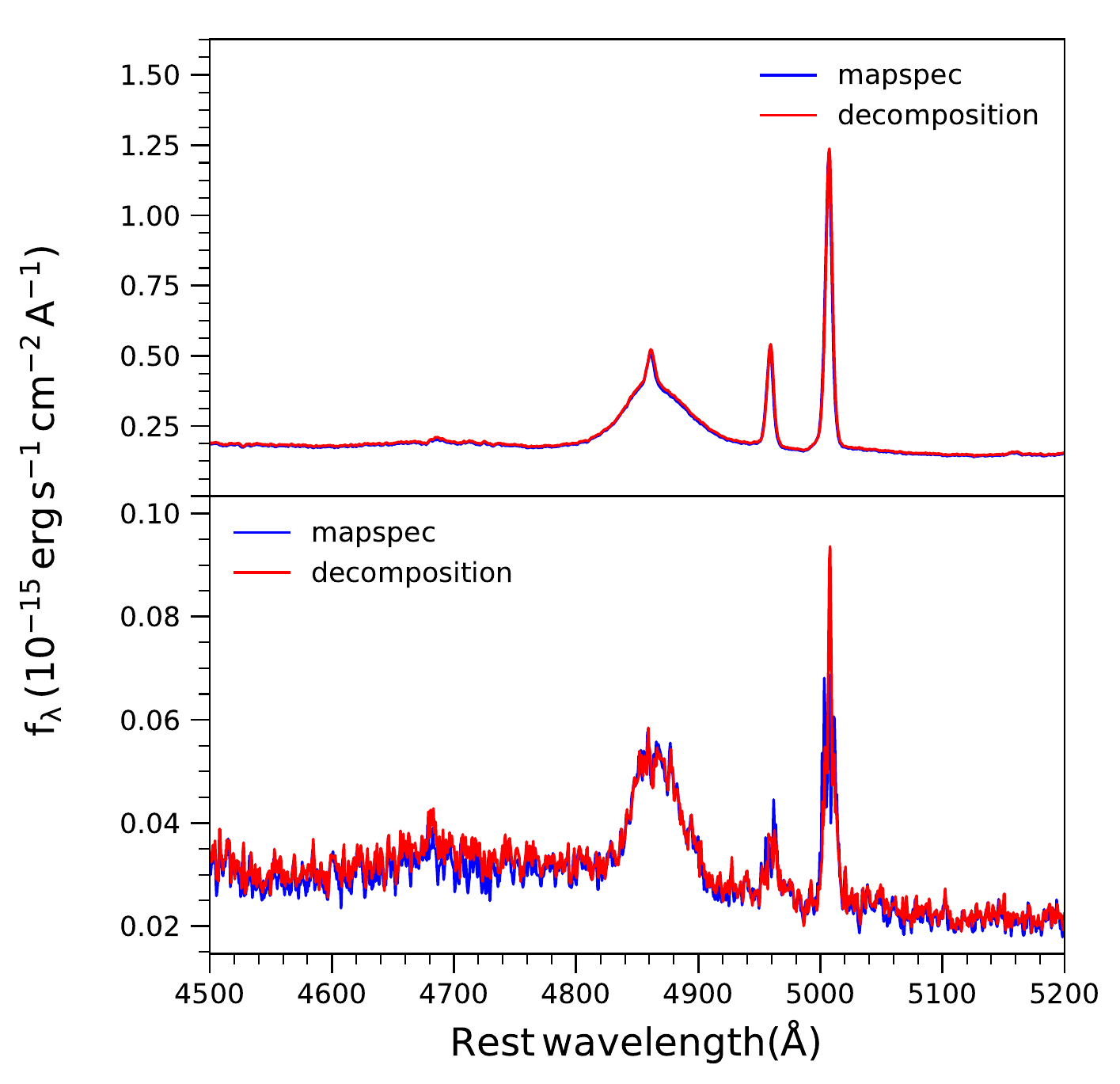}}
\resizebox{9cm}{7cm}{\includegraphics{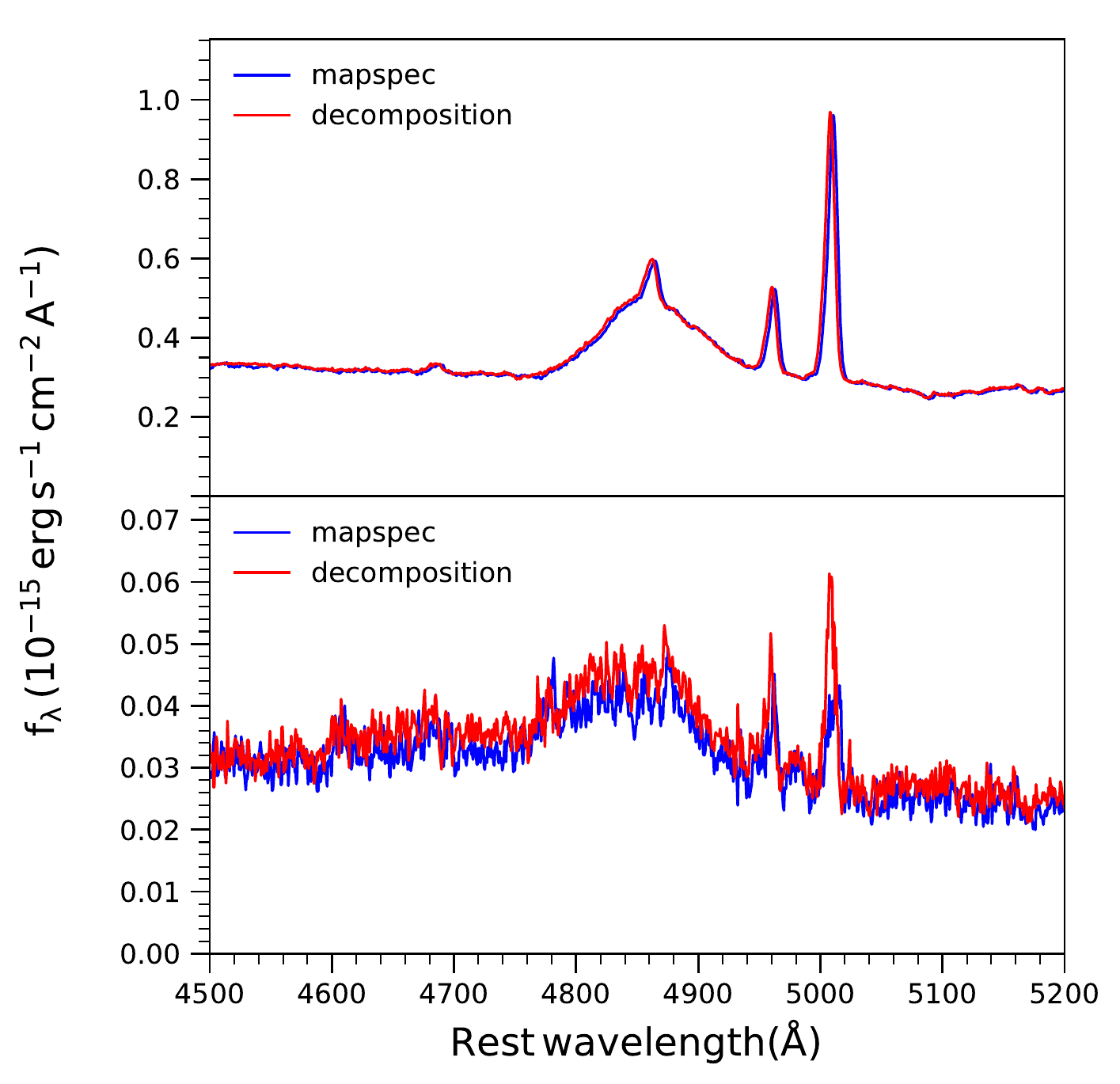}}
\caption{Mean and rms spectra of 2MASS J1026 (top) and SDSS J1619 (bottom) calculated with the mapspec code. The mean and rms spectra (red) as obtained by our spectral decomposition (see Figure \ref{Fig:mean_rms}) is overplotted for comparison.}\label{Fig:mean_rms_map} 
\end{figure}

\subsection{Line width}\label{sec:line_width}

To measure the width of H$\beta$, we constructed mean and rms spectra using the [O III]-rescaled single-epoch spectra as:
\begin{equation}
<f(\lambda)> = \frac{1}{N} \sum_{i=1}^{N} f_i (\lambda)
\end{equation}

\begin{equation}
\delta (\lambda) =\sqrt{\left[ \frac{1}{N-1} \sum_{i=1}^{N} [f_i (\lambda) -  <f(\lambda)>]^2 \right]}.
\end{equation}
Here, $f_i (\lambda)$ is the $i$th spectrum and the integration runs from 1 to the total number of spectra (N). 
The mean and rms spectra calculated with/without subtracting the power-law component and Fe II emission from individual spectra are shown in Figure \ref{Fig:mean_rms}. The rms spectrum generated after subtracting the continuum and Fe II emission clearly shows strong variation in the H$\beta$ line. It also shows the [O III] emission line, indicating that the flux calibration is not perfect. This is mainly due to the spectral misalignment as the individual spectra was obtained from two different telescopes and different setups. As a consistency check, we also used the {\scriptsize mapspec} code developed by \citet{2017PASP..129b4007F} to perform the flux rescaling. The code is based on the flux scaling algorithm of \citet{1992PASP..104..700V} but with several advantages such as lower dependency on the spectral resolution and a better smoothing kernel with Gauss-Hermite polynomials for changes in the spectral resolution. Using the {\scriptsize mapspec}, we rescaled the individual spectra and constructed mean and rms spectra as shown in Figure \ref{Fig:mean_rms_map}, which shows a good agreement between the two methods.

We measured the FWHM and line dispersion ($\sigma_{\mathrm{line}}$) from the mean and rms spectra, which were generated 
with power-law+Fe II subtracted spectra.
As shown in Figure \ref{Fig:mean_rms}, the continuum and Fe II subtracted mean spectrum has a zero continuum, while the rms spectrum has a non-zero continuum, which is presumably due to the photon noise along with the systematic error in the spectral decomposition. Thus, we first performed a linear fit to the continuum on either side of the line and subtracted the best-fit model continuum, and then measured the line width. We used a Monte Carlo bootstrap method \citep{2004ApJ...613..682P} to estimate the uncertainty of the line width measurements. We randomly selected N spectra from a set of N spectra without replacement and calculated the line width from the mean and rms spectra for each realization. This process was repeated for 5000 realizations, providing a distribution of FWHM and $\sigma_{\mathrm{line}}$. For line width measurements, we also randomly changed the endpoints of the integration region in each iteration within $\pm 10$ \AA \, from our initial selections. The mean and standard deviation of the distribution were taken as the line width and its uncertainty, respectively. 

Considering the effect of the narrow component of H$\beta$, we measured the line width from the total H$\beta$ line profile as well as from the broad component only (see Table \ref{Table:mean_rms}). For this test, we constructed the mean and rms spectra with/without subtracting the narrow H$\beta$ component from each spectrum. 
The $\sigma_{\mathrm{line}}$ measurements are similar with/without subtracting the narrow component of H$\beta$, as $\sigma_{\mathrm{line}}$ is insensitive to the peak of the line profile. The mean spectrum shows a higher $\sigma_{\mathrm{line}}$ than the rms spectrum for 2MASS J1026. Such a discrepancy has been noted in the previous studies \citep[e.g.,][]{2006ApJ...651..775B,2017ApJ...847..125P}. However, in the case of SDSS J1619, we found no significant difference in $\sigma_{\mathrm{line}}$. For line width measurements, we corrected for the instrumental resolution of $R=624$ of the Lick spectra, as the MDM spectra also have a similar resolution. The resolution corrections changed the line width by only $1-2$ \%. 

Note that we used different configuration and different spectrographs, the mean and rms spectra include systemic uncertainty. Thus, we also performed consistency checks using subsets of the data obtained with the same spectrograph with the same configuration.
For 2MASS J1026, we obtained a total of 45 spectra, of which 24 were from Lick (7 from the first CCD and 17 from the second CCD)
while for SDSS J1619, we obtained a total of 46 spectra, of which 32 were from Lick (12 from first CCD and 20 from the second CCD). We measured the line widths from the mean and rms spectra, which were generated from the spectra obtained with the second CCD at the Lick since this subset of the data contains the maximum number of epochs with a fixed configuration. We obtained consistent line width measurements (see Table \ref{Table:mean_rms_mdm}) compared 
to those based on the total epochs (see Table \ref{Table:mean_rms}). In addition, we used  {\scriptsize mapspec} to measure the line width based on the mean and rms spectra generated after subtracting the PL+Fe II from individual spectra (Table \ref{Table:mean_rms_map}). Again, we found that both methods 
provide consistent results.

 \begin{table}
 \caption{Rest-frame resolution corrected line width measurements from mean and rms spectra where power-law + Fe II were subtracted from individual spectra. Columns are (1) object name, (2) type of spectra, (3)-(4) line FWHM and dispersion from total H$\beta$, (5)-(6) line FWHM and dispersion from broad H$\beta$ component.}
	\begin{center}
	\hspace*{-1cm}
 	\resizebox{1.1\linewidth}{!}{%
     \begin{tabular}{ l l l l l l l}\hline \hline 
    Object & Type     &  \multicolumn{2}{c}{$\Delta V$ (BC+NC)} &  \multicolumn{2}{c}{$\Delta V$ (BC)}\\
           &      &  FWHM     &  $\sigma_{\mathrm{line}}$ &  FWHM &  $\sigma_{\mathrm{line}}$ \\
           &      & (km s$^{-1}$) & (km s$^{-1}$) & (km s$^{-1}$) & (km s$^{-1}$) \\
      (1)  &   (2)  & (3)  & (4) & (5) & (6)   \\\hline
     2MASS J1026    & mean  & $2441\pm73$   & $1452\pm68$   &  $3821\pm56$  & $1495\pm69$  \\
                    & rms   & $2562\pm263$  & $1122\pm73$   &  $2653\pm273$ & $1129\pm75$ \\
     SDSS J1619     & mean  & $3877\pm94$   & $2206\pm61$   &  $5750\pm74$  & $2262\pm62$ \\
                    & rms   & $6244\pm1209$ & $2487\pm139$  &  $6339\pm1204$& $2500\pm140$ \\ \hline            
     \hline 
        \end{tabular} } 
        \label{Table:mean_rms}
        \end{center}
    \end{table}

   \begin{table}
   \caption{Rest-frame resolution corrected line width measurements from mean and rms spectra created from a subset of Lick spectra (i.e., spectra after September 2016).}
  	\begin{center}
  	\hspace*{-1cm}
   	\resizebox{1.1\linewidth}{!}{%
       \begin{tabular}{ l l l l}\hline \hline 
      Object & Type     &  \multicolumn{2}{c}{$\Delta V$ (BC+NC)} \\
             &      &  FWHM     &  $\sigma_{\mathrm{line}}$  \\
             &      & (km s$^{-1}$) & (km s$^{-1}$) \\
        (1)  &   (2)  & (3)  & (4)   \\\hline
       2MASS J1026    & mean  & $2244 \pm 194$    & $1465\pm 67$   \\
                      & rms   & $3011 \pm 620$   & $1171\pm 190 $  \\
       SDSS J1619     & mean  & $4017 \pm 184$    & $2262\pm 62 $   \\
                      & rms   & $5355 \pm 827$  & $2500\pm 141 $  \\ \hline            
       \hline 
          \end{tabular} } 
          \label{Table:mean_rms_mdm}
          \end{center}
      \end{table}

   \begin{table}
   \caption{Rest-frame resolution corrected line width measurements from mean and rms spectra obtained by {\scriptsize mapspec} code where power-law + Fe II were subtracted from individual spectra.}
  	\begin{center}
  	\hspace*{-1cm}
   	\resizebox{1.1\linewidth}{!}{%
       \begin{tabular}{ l l l l}\hline \hline 
      Object & Type     &  \multicolumn{2}{c}{$\Delta V$ (BC+NC)} \\
             &      &  FWHM     &  $\sigma_{\mathrm{line}}$  \\
             &      & (km s$^{-1}$) & (km s$^{-1}$) \\
        (1)  &   (2)  & (3)  & (4)   \\\hline
       2MASS J1026    & mean  & $2456 \pm 72$    & $1455\pm 68$   \\
                      & rms   & $2656 \pm 267$   & $1178\pm 74 $  \\
       SDSS J1619     & mean  & $3894 \pm 80$    & $2189\pm 68 $   \\
                      & rms   & $5551 \pm 1063$  & $2389\pm 163 $  \\ \hline            
       \hline 
          \end{tabular} } 
          \label{Table:mean_rms_map}
          \end{center}
      \end{table}

 \begin{table}
 \caption{Black hole mass measurements. Columns are (1) object name, (2) type of spectra, (3) types of line width, (4) black hole mass from total H$\beta$, (5) black hole mass from broad H$\beta$ component.}
	\begin{center}
	\hspace*{-1cm}
 	\resizebox{1.1\linewidth}{!}{%
     \begin{tabular}{ l l l l l}\hline \hline 
    Object                      & Spectrum    & $\Delta V$      &  $M_{\mathrm{BH}}$ (BC+NC) &  $M_{\mathrm{BH}}$ (BC)\\
                                &             &                 & ($\times 10^7 \, M_{\odot}$)  &  ($\times 10^7 \, M_{\odot}$) \\
      (1)                       &   (2)       &   (3)           &        (4)        &       (5)   \\\hline
     2MASS J1026    & Mean  & FWHM       								 & $4.32^{+0.52}_{-0.63}$    & $10.61^{+1.25}_{-1.53}$  \\
     							&       & $\sigma_{\mathrm{line}}$       & $6.11^{+0.77}_{-0.92}$    & $6.48^{+0.81}_{-0.97}$   \\
                                & rms   & FWHM                           & $4.76^{+0.74}_{-0.84}$    & $5.11^{+0.79}_{-0.90}$   \\
                                &       & $\sigma_{\mathrm{line}}$       & $\mathbf{3.65^{+0.49}_{-0.57}}$    & $3.69^{+0.49}_{-0.58}$  \\
                                
    SDSS J1619    & Mean  & FWHM                                         & $14.02^{+4.70}_{-3.93}$   & $30.84^{+10.32}_{-8.63}$   \\
         					    &       & $\sigma_{\mathrm{line}}$       & $18.12^{+6.08}_{-5.09}$   & $19.04^{+6.39}_{-5.34}$  \\
                & rms   & FWHM                                           & $36.37^{+14.05}_{-12.35}$ & $37.49^{+14.41}_{-12.65}$   \\
                                &       & $\sigma_{\mathrm{line}}$       & $\mathbf{23.02^{+7.81}_{-6.56}}$   & $23.26^{+7.89}_{-6.62}$ \\
     \hline 
        \end{tabular} } 
        \label{Table:bh_mass}
        \end{center}
    \end{table}

\subsection{Black hole mass}\label{sec:BH_mass}

By combining $R_{\mathrm{BLR}}$ calculated from the rest-frame $\tau_{\mathrm{cent}}$ and line width of H$\beta$, we determined black hole mass using equation \ref{eq:mass}.
For the line width, we used either $\sigma_{\mathrm{line}}$ or FWHM of H$\beta$, which were measured from the mean or rms spectra.
As the mean and rms spectra were constructed using individual H$\beta$ line profiles with/without subtracting the narrow component of H$\beta$, in total, we have 8 different choices for the line widths (see Table \ref{Table:mean_rms}).
In the case of the virial factor, we adopted $f = 4.47$ and 1.12, respectively for the line dispersion and FWHM of H$\beta$ as determined by \citet{2015ApJ...801...38W}. Adopting those $f$ values, we determined black hole masses, and their uncertainties were estimated via error propagation. 

For 2MASS J1026, using the rest-frame H$\beta$ $\tau_{\mathrm{cent}}$=$33.2^{+3.9}_{-4.7}$ days, we determined black holes mass as $4-10\times10^7 M_{\odot}$  depending on the choice of the line width (see Table \ref{Table:bh_mass}). In the case of SDSS J1619, we used the H$\beta$ $\tau_{\mathrm{cent}}$= $42.6^{+14.3}_{-11.9}$ days, obtaining black hole mass as $1-3\times 10^8 M_{\odot}$ (see Table \ref{Table:bh_mass}). Note that the uncertainty of black hole mass was estimated by propagating measurement errors. 
The $\sigma_{\mathrm{line}}$ based on the rms spectrum is widely used for black hole mass estimation, since  $\sigma_{\mathrm{line}}$ is less sensitive to the peak of the line, while the FWHM is an alternative choice since it is less sensitive to the wing
\citep[e.g.,][]{2004ApJ...613..682P}. We present the black hole mass based on $\sigma_{\mathrm{line}}$ measured from the total H$\beta$ profile in the rms spectrum as the best measurement. Note that the subtraction of the narrow component of H$\beta$ is negligible
in determining black hole mass (see Table \ref{Table:bh_mass}), while the different choice of the line width changes black hole mass by a factor of $\sim$2.

 \begin{figure}
 \centering
 \resizebox{9cm}{7.5cm}{\includegraphics{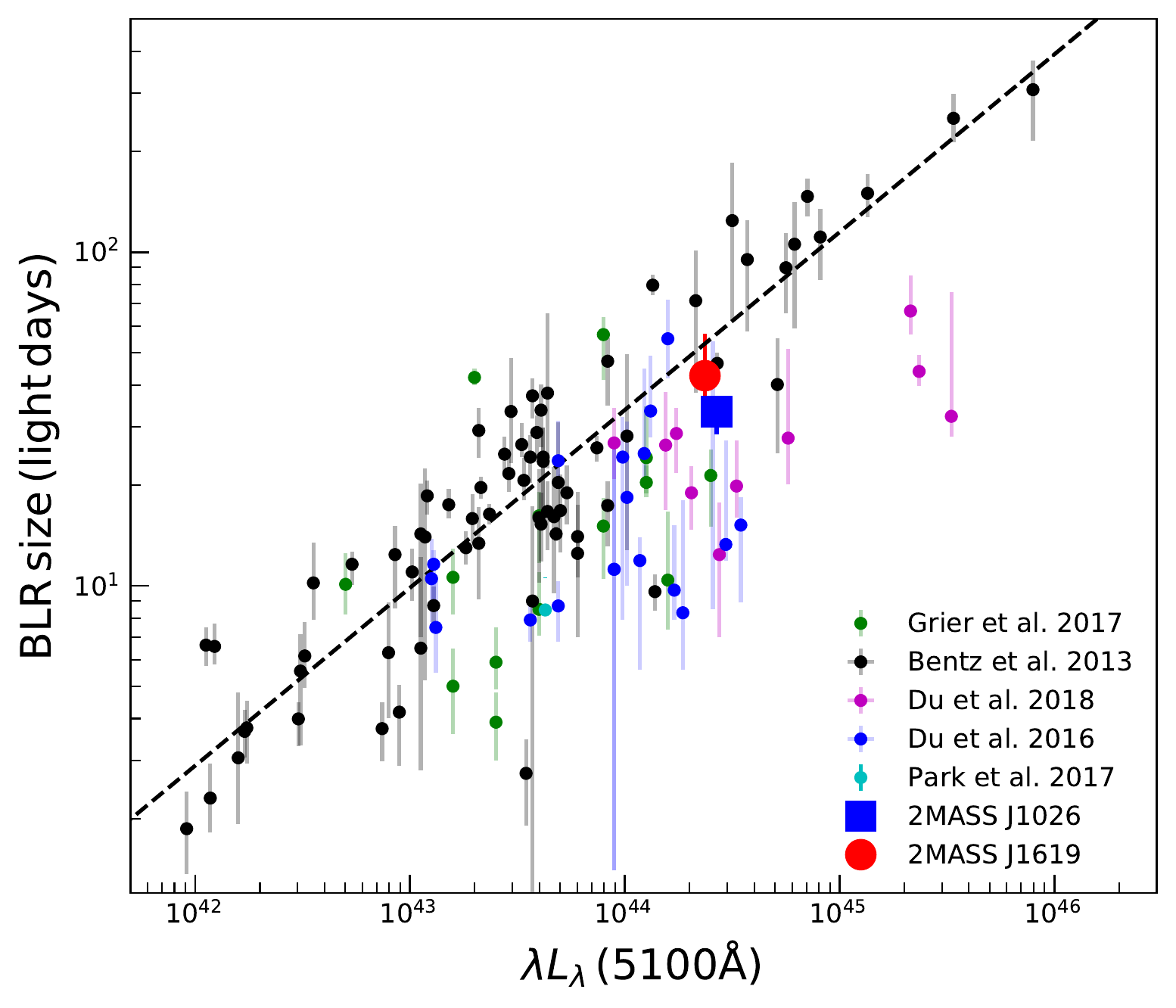}}
 \caption{Size-luminosity relation. The best fitted relation (dashed line) of \citet{2013ApJ...767..149B} and the various samples in the literature 
 are presented. From \citet{2017ApJ...851...21G}, only objects with best quality lag measurement (quality rating of 5) are shown.}\label{Fig:lag_lum} 
 \end{figure}


\section{Discussion}\label{sec:discussion}

\subsection{Systematic uncertainty of the lag measurement}

We selected the $B$-band for tracing AGN continuum variability, in order to exclude the effect of the H$\beta$ emission line for 
the two targets at $z\sim 2.4-2.6$. On the other hand, $B$-band may not be the best choice since it is centered at the Balmer jump (3646\AA), which may contribute to the disk continuum emission \citep[e.g.][]{2018ApJ...857...53C, 2019ApJ...870..123E}. 
In addition, the high-order Balmer lines can contribute to the total flux observed with the $B$-band. Thus, the H$\beta$ lags relative to the $B$-band photometry could be underestimated. To test this effect, we performed the cross correlation analysis using the 5100\AA \ flux light curve instead of the $B$-band photometry light curve, and measured the H$\beta$ time lag (see section \ref{sec:lag}). Out of the two AGNs, only  2MASS J1026 showed a good quality light curve based on 5100\AA\ flux, and we measured the H$\beta$ time lag relative to $f_{5100}$ as $43.4^{+6.4}_{-7.5}$ days (see Table \ref{Table:ccf}), which is consistent with the $B$-band based lag within the error, suggesting that the effect of the Balmer jump is negligible for the target.

\subsection{Mass measurement}

In this section, we compare our reverberation-based black hole mass with the previously reported single-epoch black hole masses. 
\citet{2011ApJS..194...45S} reported $\log M_{\mathrm{BH}} = 8.09\pm0.02 M_{\odot}$ for 2MASS J1026 and $8.38\pm0.01 M_{\odot}$ for SDSS J1619, using SDSS DR7 single-epoch spectra. Note that single-epoch black hole mass varies significantly depending on the choice of the BLR size-luminosity relation, the method of line width measurement, and the choice of the scale factor. To make a consistent comparison, we calculated single-epoch mass using the monochromatic luminosity and line width measurements from \citet{2011ApJS..194...45S}, 
and the updated BLR size-luminosity relation of \citet{2013ApJ...767..149B}.
Using the FWHM from the single-epoch SDSS spectra and a $f=1.12$, we obtained $M_{\mathrm{BH}}=1.1 \times 10^8 M_{\odot}$ ($2.2 \times 10^8 M_{\odot}$) for 2MASS J1026 (SDSS J1619).  Thus, the reverberation-based mass is a factor of 3 lower than the single-epoch mass for 2MASS J1026, while the two black hole masses are consistent in the case of SDSS J1619.  
As the scatter of the BLR size-luminosity relation is much larger than 0.3 dex based on recent studies \citep[e.g.,][]{Pei_2017}, the single-epoch mass suffers large uncertainties, while the indirect single-epoch mass is still useful for studying a population of AGNs. 

\subsection{Size-luminosity relation}\label{sec:Size-luminosity}

We investigate the H$\beta$ BLR size-luminosity by adding the two new lag measurements. 
We measured the monochromatic luminosity at 5100 \AA, ($L_{5100}$) from the best-fit power-law component using the mean spectrum as described in section \ref{sec:spectra_analysis}. Considering the uncertainty in the flux calibration, we rescaled the mean spectrum based on the photometry. Since the mean spectrum do not cover the $B$-band, we used the mean $V$-band magnitude of $17.80\pm0.23$ and $17.58\pm0.12$ respectively for 2MASS J1026 and SDSS J1619 from our photometry monitoring. In this process, we obtained a scale factor of 1.35 and 1.07 for 2MASS J1026 and SDSS J1619, respectively, in order to match the synthetic $V$-band magnitude of the mean spectrum to the mean $V$-band magnitude. Based on the re-scaled mean spectrum, we measured $L_{5100}$ of $2.67\times10^{44}$ erg s$^{-1}$ for 2MASS J1026 and $2.87 \times 10^{44}$ erg s$^{-1}$ for SDSS J1619. 

Based on the size-luminosity relation of \citet{2013ApJ...767..149B}, we calculated the expected lag from the measured $L_{5100}$
as $\sim 57$  and $\sim59$ light days, respectively for 2MASS J1026 and SDSS J1619, which are slightly larger than the reverberation-mapped BLR size. Considering the scatter of the size-luminosity relation by \citet{2013ApJ...767..149B}, we found that the two AGNs follow the relation without a significant offset. 

We considered the effect of the host galaxy contribution to the measured AGN luminosity. Since the spectra were taken at different position
angles and various seeing conditions, a different part of the host galaxy was observed through the slit. Thus, the spectroscopic continuum variability can be affected by the host-galaxy contribution. This issue is potentially important for SDSS J1619, which shows a slightly extended host galaxy structure in the SDSS image, while 2MASS J1026 appears to be a point source. To quantify the AGN fraction, we stacked $B$-band and $V$-band images of SDSS J1619 obtained at the MDM 1.3m telescope, in order to measure the host galaxy fraction. Using GALFIT \citep{2002AJ....124..266P}, we performed host galaxy decomposition, and obtained host galaxy fraction as $\sim$8 and 18\% of the total flux in the $B$ and $V$-band, respectively. 
As a consistency check, we also measured the host galaxy contribution in the mean spectrum of SDSS J1619 by performing spectral decomposition (see section \ref{sec:analysis}). By excluding AGN emission, i.e., a power-law component, Fe II emission blends, and emission lines, we determined the host galaxy flux from the stellar component, which was modeled with a stellar template constructed based on the Indo-US spectral library \citep{2004ApJS..152..251V}. The host galaxy fraction is found to be 28\% in the $V$-band, which is similar to that obtained from the $V$-band image decomposition. To correct for the host galaxy contribution, we adopted the host galaxy fraction from the photometry decomposition, and reduced the continuum luminosity by 18\% as $L_{5100}$ = $2.35 \times 10^{44}$ erg s$^{-1}$ for SDSS J1619.

In Figure \ref{Fig:lag_lum}, we present the BLR size-luminosity relation 
by adding various recent results \citep{2016ApJ...825..126D,2018ApJ...856....6D, 2017ApJ...851...21G, 2017ApJ...847..125P}. Our measured lags and  the host galaxy-corrected $L_{5100}$ are consistent with the best-fit relation of \citet{2013ApJ...767..149B}.
Note that the monochromatic luminosity at 5100\AA\ of new AGNs, which are added to the sample of \citet{2013ApJ...767..149B}, suffers large uncertainty due to the lack of the careful imaging analysis with high spatial resolution. High-resolution imaging data are required to investigate whether the continuum luminosity of these AGNs is overestimated due to the contribution of the host galaxy, and to accurately measure $L_{5100}$. 

Recent reverberation campaigns of high Eddington ratio AGNs show a significant offset from the BLR size-luminosity relation
\citep{2016ApJ...825..126D}. High accretion AGNs may have slim accretion disks, producing strong self-shadowing effects leading to the highly anisotropic radiation field and two dynamically distinct regions in the BLR \citep{Wang_2014}. Such a scenario has been supported by the geometrical and kinematical modeling of Mrk 142 \citep{2018ApJ...869..137L}, suggesting that the radiation field and BLR geometry may be more complex in highly mass-accreting AGNs.  

We calculated the Eddington ratio ($\lambda_{\mathrm{EDD}}$) 
using $L_{\mathrm{EDD}}=1.26\times10^{38} \, M_{\mathrm{BH}}$, while the bolometric luminosity was estimated from the scaled $L_{5100}$ using $L_{\mathrm{BOL}}=9\times L_{5100}$ \citep{2000ApJ...533..631K}. Our AGNs are not super-Eddington sources as 
$\lambda_{\mathrm{EDD}}$ are $\sim0.52$ and 0.07, respectively for 2MASS J1026 and SDSS J1619. Thus, the two AGNs show
no clear offset from the size-luminosity relation. 

The deviation from the BLR size-luminosity relation is also noticed in low accreting sources e.g., NGC 5548 \citep[$\lambda_{\mathrm{EDD}}=0.021$,][]{2010MNRAS.402.1081V}. The H$\beta$ lag of NGC 5548 is 5 times smaller than the expected result based on the BLR size-luminosity relation \citep{Pei_2017}. A change in the response of the emission lines to the ionization continuum was noticed during the monitoring campaign, suggesting complex behaviors of the ionization radiation field. Repeated observations of more individual AGNs will help to understand the size-luminosity relation in detail. Our ongoing campaign for a larger number of AGNs may shed light on the scatter and the validity of the size-luminosity relation, especially at the high luminosity end.      

\section{Summary}\label{sec:summary}

We present the first reverberation mapping results from the SNU AGN Monitoring Project. Based on the first three year campaign, we obtained the $B$-band and H$\beta$ light curves, determining the H$\beta$ BLR size for two objects, 2MASS J1026 and SDSS J1619. Our main results are summarized as follows.

\begin{itemize}
\item Based on cross-correlation analysis using $B$-band and H$\beta$ emission line light curves, we measured the rest-frame lag of $33.2^{+3.9}_{-4.7}$ and $42.6^{+14.3}_{-11.9}$ days, respectively for 2MASS J1026 and SDSS J1619. 


\item  Using the $\sigma_{\mathrm{line}}$ of the total H$\beta$ line profile in the rms spectrum and a virial factor of $f=4.47$, we determined black hole mass as $3.65^{+0.49}_{-0.57} \times 10^7 M_{\odot}$ ($23.02^{+7.81}_{-6.56} \times 10^7 M_{\odot}$) for 2MASS J1026 (SDSS J1619). 

\item 
Within the uncertainty of the monochromatic luminosity at 5100\AA, the two AGNs show a consistent BLR size-luminosity relation compared to the sample of \citet{2013ApJ...767..149B}.
\end{itemize} 
 
\acknowledgments
We thank the referee for useful suggestions, which improved the clarity of the manuscript.
This work has been supported by the Basic Science Research Program through the National Research Foundation of Korea government (2016R1A2B3011457), and by Samsung Science and Technology Foundation under Project Number SSTF -BA1501-05.
This work is based on observations obtained at the MDM Observatory, operated by Dartmouth College, Columbia University, Ohio State University, Ohio University, and the University of Michigan. We thank the Lick Observatory staff for their enormous help during our observations. We also thank the operators at LOAO, Jaehyuk Yoon and Inkyung Baek for executing our program. SR thanks Neha Sharma (KHU) for carefully reading the manuscript.  

 \bibliographystyle{apj}
 \bibliography{ref}

\end{document}